\documentclass[twocolumn,showpacs,superscriptaddress,preprintnumbers,prd]{revtex4}
\usepackage{graphicx,amsmath,dcolumn,bm}
\begin{document}
\preprint{KEK-TH-1013}
\title{\Large \bf Determination of nuclear parton distribution functions \\
                  and their uncertainties at next-to-leading order}
\author{M. Hirai}
\affiliation{Department of Physics,
             Tokyo Institute of Technology,
             Ookayama, Meguro-ku, Tokyo, 152-8550, Japan}
\affiliation{Institute of Particle and Nuclear Studies,
          High Energy Accelerator Research Organization (KEK) \\
          1-1, Ooho, Tsukuba, Ibaraki, 305-0801, Japan}
\author{S. Kumano}
\affiliation{Institute of Particle and Nuclear Studies,
          High Energy Accelerator Research Organization (KEK) \\
          1-1, Ooho, Tsukuba, Ibaraki, 305-0801, Japan}
\affiliation{Department of Particle and Nuclear Studies,
             Graduate University for Advanced Studies \\
           1-1, Ooho, Tsukuba, Ibaraki, 305-0801, Japan}     
\author{T.-H. Nagai}
\affiliation{Department of Particle and Nuclear Studies,
            Graduate University for Advanced Studies \\
           1-1, Ooho, Tsukuba, Ibaraki, 305-0801, Japan}      
\homepage[URL: ]{http://research.kek.jp/people/kumanos/nuclp.html}
\date{November 7, 2007}
\begin{abstract}
Nuclear parton distribution functions (NPDFs) are determined by global
analyses of experimental data on structure-function ratios $F_2^A/F_2^{A'}$
and Drell-Yan cross-section ratios $\sigma_{DY}^A/\sigma_{DY}^{A'}$.
The analyses are done in the leading order (LO) and next-to-leading order (NLO)
of running coupling constant $\alpha_s$. Uncertainties of the NPDFs are
estimated in both LO and NLO for finding possible NLO improvement.
Valence-quark distributions are well determined, and antiquark distributions
are also determined at $x<0.1$. However, the antiquark distributions have
large uncertainties at $x>0.2$. Gluon modifications cannot be fixed
at this stage. Although the advantage of the NLO analysis, in comparison
with the LO one, is generally the sensitivity to the gluon distributions,
gluon uncertainties are almost the same in the LO and NLO. It is because
current scaling-violation data are not accurate enough to 
determine precise nuclear gluon distributions. Modifications of
the PDFs in the deuteron are also discussed by including data
on the proton-deuteron ratio $F_2^D/F_2^p$ in the analysis.
A code is provided for calculating the NPDFs and their uncertainties
at given $x$ and $Q^2$ in the LO and NLO.
\end{abstract}

\pacs{13.60.Hb, 12.38.-t, 24.85.+p, 25.30.-c}
\maketitle

\section{Introduction}

Cross sections of high-energy nuclear reactions are expressed in terms of
nuclear parton distribution functions (NPDFs), so that precise NPDFs
are essential for finding any new phenomena in the high-energy reactions.
Recently, this topic is becoming important in heavy-ion collisions
for investigating properties of quark-hadron matters \cite{rhic-heavy-ion}
and also in neutrino reactions for investigating neutrino-oscillation
physics \cite{nuint}. 
Determination of precise NPDFs is valuable for studying various
phenomena in heavy-ion reactions such as color glass condensate
\cite{cgc-summary}, $J/\psi$ suppression
\cite{at06-j-psi}, and parton-energy loss \cite{heavy-ion-renk-etal}.
The NPDF studies should be also important for heavy-ion collisions 
at LHC (Large Hadron Collider) \cite{lhc-pdf-2004}.

In neutrino oscillation experiments, most data are taken
at small $Q^2$ ($<$1 GeV$^2$). We could approach such
a kinematical region from the high-energy deep inelastic one
by using quark-hadron duality \cite{duality}.
However, there are still unresolved issues
in neutrino deep inelastic scattering. For example, an anomalous
$\sin^2 \theta_W$ value was reported in the neutrino-iron scattering
by the NuTeV collaboration \cite{nutev02}. It could be related to
a nuclear modification difference between the parton distributions
$u_v$ and $d_v$ \cite{npdf-sinth,brodsky-nu-04,ep06-sinth}
because the iron target is used in the NuTeV measurements.
There is also an issue that nuclear corrections are
different from the ones expected from electron and muon scattering 
experiments according to recent NuTeV data \cite{cteq07-small-A-corr}.

In these high-energy nuclear reactions, nucleonic PDFs rather than
the nuclear ones are often used in calculating cross sections by neglecting
nuclear modifications although it is well known that nuclear corrections
could be as large as 20\% in medium-size nuclei \cite{sumemc}. 
These nuclear modifications have been experimentally investigated
mainly by the measurements of structure-function ratios $F_2^A/F_2^{A'}$
and Drell-Yan cross-section ratios $\sigma_{DY}^A/\sigma_{DY}^{A'}$. 
Physical mechanisms of the nuclear corrections are, for example,
summarized in Ref. \cite{sumemc}. In the small-$x$ region, the NPDFs
become smaller than the corresponding nucleonic ones, which is called
shadowing. There are depletions at medium $x$, which is related
to the nuclear binding mechanism and possibly to a nucleonic modification
inside a nuclear medium \cite{nucl-corr}. At large $x$, the nucleon's
Fermi motion gives rise to positive corrections. 

Because the PDFs are related to the nonperturbative aspect of quantum
chromodynamics (QCD), theoretical calculations have been done by lattice QCD
or phenomenological models. However, such calculations are not accurate
enough at this stage. One would like to have accurate NPDFs,
which are obtained in a model-independent way,
for calculating precise nuclear cross sections. We should
inevitably rely on experimental data for determining them.

Studies of nucleonic PDFs have a long history with abundant
experimental data in a wide kinematical region \cite{durham}. 
However, determination of NPDFs is still at a premature
stage with the following reasons. First, available experimental data are
limited. The experiments of the Hadron-Electron Ring Accelerator (HERA)
provided data for structure functions at small $x$ in a wide range
of $Q^2$; however, such data do not exist for nuclei. Because of
final-state interactions, hadron-production data may not be suitable
for the NPDF determination, whereas they are used in the nucleonic analysis.
Second, the analysis technique is not established. Parametrization studies
for the NPDFs started only recently. The NPDFs are expressed in terms of
a number of parameters which are then determined by a $\chi^2$ analysis
of the nuclear data. However, it is not straightforward to find
functional forms of mass-number ($A$) and Bjorken-$x$ dependencies
in the NPDFs. Furthermore, higher-twist effects could be important
in the small-$Q^2$ region.

A useful parametrization was investigated in Ref. \cite{eskoka-etal}
by analyzing $F_2$ structure functions and Drell-Yan data; however,
the distributions were obtained by simply assigning appropriate 
parameter values by hand in the versions of 1998 and 1999.
The first $\chi^2$ analysis was reported in Ref. \cite{hkm01},
and then uncertainties of the NPDFs were obtained \cite{hkn04}.
All of these analyses are done in the leading order (LO) of
the running coupling constant $\alpha_s$. A next-to-leading-order (NLO)
analysis was recently reported \cite{ds04}. The LO $\chi^2$ analysis
with the uncertainties was also investigated in the 2007 version
of Ref. \cite{eskoka-etal}. There are related studies on the nuclear
shadowing \cite{armesto-shadow,qiu-shadow,fgs05-shadow}
and a global analysis of structure functions \cite{kp06}. 

In this way, the parametrization studies have been developed
recently for the NPDFs, and they are not sufficient.
Here, we extend our studies in Refs. \cite{hkm01,hkn04} by focusing
on the following points:
\begin{itemize}
\item NLO analysis with NPDF uncertainties together with a LO one,
\item roles of NLO terms on the NPDF determination 
      by comparing LO and NLO results,
\item better determination of $x$ and $A$ dependence,
\item nuclear modifications in the deuteron
      by including $F_2(\text{deuteron})/F_2(\text{proton})$ data,
\item flavor asymmetric antiquark distributions.
\end{itemize}

This article is organized as follows. In Sec. \ref{analysis}, our analysis
method is described for determining the NPDFs. Analysis results
are explained in Sec. \ref{results}. Nuclear modifications in the deuteron
are discussed in Sec. \ref{deuteron}. The results are summarized in
Sec. \ref{summary}.

\section{\label{analysis} Analysis method}

The optimum NPDFs are determined by analyzing experimental data
of the $F_2$ structure functions and Drell-Yan cross sections
for nuclear targets. Details of our analysis method are described
in Refs. \cite{hkm01,hkn04}, so that only the outline is explained
in the following.

\subsection{\label{paramet} Parametrization}

The parton distribution functions are expressed by two variables $x$ and
$Q^2$. The variable $Q^2$ is defined $Q^2=-q^2$ with the virtual photon
momentum $q$ in the lepton scattering, and the scaling variable $x$ is
given by $x=Q^2/(2M\nu)$ with the nucleon mass $M$ and the energy
transfer $\nu$. The variables for the Drell-Yan process are
momentum fractions, $x_1$ and $x_2$ for partons
in the projectile and target, respectively, and $Q^2$ defined by
the dimuon mass as $Q^2=m_{\mu\mu}^2$.
In our analysis, the NPDFs are expressed in terms of corresponding
nucleonic PDFs multiplied by weight functions:
\begin{equation}
f_i^A (x, Q_0^2) = w_i(x,A,Z) \, f_i (x, Q_0^2).
\label{eqn:paramet}
\end{equation}
The functions $f_i^A$ and $f_i$ indicate type-$i$ NPDF and nucleonic PDF,
respectively, and $w_i$ is a weight function which indicates a nuclear
modification for the type-$i$ parton distribution. The function $w_i$
generally depends on not only $x$ and $A$ but also the atomic number $Z$.
It should be noted that this expression sacrifices the large-$x$ ($>1$)
nuclear distributions. Finite distributions exist even at $x>1$
in nuclei, whereas the distributions should vanish in the nucleon.

Flavor symmetric antiquark distributions are assumed for $\bar u$, $\bar d$,
and $\bar s$ in the previous analyses \cite{hkm01,hkn04}. From the violation
of the Gottfried sum rule and Drell-Yan measurements, it is now well known
that these antiquark distributions are different \cite{flavor3}. 
It is more natural to investigate modifications from the flavor asymmetric
antiquark distributions in the nucleon. In this work, flavor asymmetric
antiquark distributions are used in nuclei, and the distribution type $i$
indicates $u_v$, $d_v$, $\bar u$, $\bar d$, $\bar s$, and $g$: 
\begin{align}
u_v^A (x,Q_0^2) & = w_{u_v} (x,A,Z) \, \frac{Z u_v (x,Q_0^2) 
                                         + N d_v (x,Q_0^2)}{A},
\nonumber \\
d_v^A (x,Q_0^2) & = w_{d_v} (x,A,Z) \, \frac{Z d_v (x,Q_0^2) 
                                         + N u_v (x,Q_0^2)}{A},
\nonumber \\
\bar u^A (x,Q_0^2) & = w_{\bar q} (x,A,Z) \, \frac{Z \bar u (x,Q_0^2) 
                                         + N \bar d (x,Q_0^2)}{A},
\nonumber \\
\bar d^A (x,Q_0^2) & = w_{\bar q} (x,A,Z) \, \frac{Z \bar d (x,Q_0^2) 
                                         + N \bar u (x,Q_0^2)}{A},
\nonumber \\
\bar s^A (x,Q_0^2) & = w_{\bar q} (x,A,Z) \, \bar s (x,Q_0^2), 
\nonumber \\
g^A (x,Q_0^2)      & = w_{g} (x,A,Z) \, g (x,Q_0^2),
\label{eqn:wpart}
\end{align}
whereas the relation $\bar u^A=\bar d^A=\bar s^A$ is assumed in
Refs. \cite{hkm01,hkn04}.
The number of flavor is three (four) at $Q^2 < m_c^2$ ($Q^2 > m_c^2$)
\cite{hkn04}. The bottom and top quark distributions are neglected.
The strange-quark distributions are assumed to
be symmetric ($s=\bar s$) although there are recent studies
on possible asymmetry $s \ne \bar s$ \cite{s-sbar}.
The charm-quark distributions are
created by $Q^2$ evolution effects \cite{hkn04}. As for the nucleonic
PDFs in the LO and NLO, the MRST (Martin, Roberts, Stirling, and Thorne)
parametrization of 1998 is used \cite{mrst98} in this analysis,
where the charm-quark mass is $m_c$=1.35 GeV and scale parameters
are  $\Lambda_4$=0.174 and 0.300 GeV for the LO and NLO, respectively.
In our previous analysis \cite{hkn04}, the MRST-2001 version
was employed. Since the NLO gluon distribution is negative
at $x<0.005$ and $Q^2=$1 GeV$^2$ in the 2001 version,
we use the 1998 parametrization in this work.
Negative gluon distributions in nuclei could affect our analysis
inappropriately in the shadowing region. Furthermore, other
researchers may use our NPDFs at small $x$ ($<0.005$) with
$Q^2 \sim$1 GeV$^2$ for calculating cross sections, for example,
in LHC experiments. We tested various PDFs of the nucleon, but
overall results are not significantly changed. Since we are
interested in obtaining the distributions at $Q^2 \ge$1 GeV$^2$
in comparison with other distributions and also our previous
distributions, we decided to use the MRST distributions. 
In the analyses of the CTEQ (Coordinated Theoretical/Experimental 
Project on QCD Phenomenology and Tests of the Standard Model) 
collaboration, the initial scale $Q^2$=(1.3)$^2$ GeV$^2$ is used.

The nuclear modification is assumed to have the following functional form:
\begin{equation}
w_i(x,A,Z)=1+ \left( 1 - \frac{1}{A^\alpha} \right) 
          \frac{a_i +b_i x+c_i x^2 +d_i x^3}{(1-x)^{\beta_i}},
\label{eqn:wi}
\end{equation}
where $\alpha$, $a_i$, $b_i$, $c_i$, $d_i$, and $\beta_i$ are parameters.
The parameter $a_i$ controls the shadowing part, $b_i$, $c_i$, and 
$d_i$ determine a minute functional form, and $\beta_i$ is related to
the Fermi-motion part at large $x$. The parameter $\beta_i$ is
fixed at $\beta_i$=0.1 because it cannot be determined from a small
number of data in the Fermi-motion part.
As it will be shown in the result section, the antiquark and gluon 
modifications cannot be determined from the present data at $x>0.3$.
If a large $\beta_i$, for example $\beta_i$=1, is taken, the antiquark
and gluon distributions could become unrealistically large at large $x$.
In order to avoid such an issue, $\beta_{i}=0.1$ is used.
The overall $A$ dependence in Eq. (\ref{eqn:wi}) is taken $\alpha=1/3$ 
simply by considering nuclear volume and surface contributions \cite{a-dep}.
There are three constraints for the parameters by the nuclear charge $Z$,
baryon number $A$, and momentum conservation \cite{hkm01,hkn04,fsl-1990}:
\begin{align}
Z &  = \int dx \, \frac{A}{3} 
             \left [ 2 \, u_v^A (x,Q_0^2) - d_v^A (x,Q_0^2) \right ],
\nonumber \\
A & = \int dx  \, \frac{A}{3} 
             \left [  u_v^A (x,Q_0^2) + d_v^A (x,Q_0^2) \right ],
\label{eqn:3conserv}
\\
A & = \int dx \, A \, x 
           \big [  u_v^A (x,Q_0^2) + d_v^A (x,Q_0^2)
\nonumber \\
  &   + 2 \, \big \{ \bar u^A(x,Q_0^2) + \bar d^A(x,Q_0^2) 
                    + \bar s^A(x,Q_0^2) \big \} + g^A (x,Q_0^2) \big ].
\nonumber
\end{align}
We selected three parameters, $a_{u_v}$, $a_{d_v}$, and $a_g$, which are
determined by these conditions.

Following improvements are made from the previous analysis 
\cite{hkn04}. First, the parametrization of $x$ dependence is modified.
The meaning of the parameters $b_i$, $c_i$, and $d_i$ is not obvious, so
that it is difficult to limit the ranges of these parameters in the analysis.
Here, we take $x$ points ($x_{0i}^+$, $x_{0i}^-$) of
extreme values for the function
$a_i +b_i x+c_i x^2 +d_i x^3$ as the parameters instead of $b_i$ and
$c_i$. They are related with each other by
\begin{equation}
b_i = 3 d_i \, x_{0i}^+ x_{0i}^-, \ \ \ 
c_i = - \frac{3d_i}{2} (x_{0i}^+ + x_{0i}^-) .
\label{eqn:bici}
\end{equation}
Then, the values of $x_{0i}^+$ and $x_{0i}^-$ become transparent
at least for the valence-quark distributions. From the measurements
of the ratios $F_2^A/F_2^D$, where $D$ indicates the deuteron,
we have a rough idea that the extreme
values should be $x_{0v}^- \sim 0.15$ and $x_{0v}^+ \sim 0.6$. 
The Drell-Yan measurements indicate $x_{0\bar q}^- \sim 0.1$ 
for the antiquark distributions. The value $x_{0\bar q}^+$
and the extreme values for the gluon distribution are not obvious. 

Second, the $A$ dependence $1/A^{1/3}$ is too simple. From gross nuclear
properties, the leading $A$ dependence could be described by
$1 - 1/A^{1/3}$. In order to describe more details, the parameters
$d_v$, $a_{\bar q}$, $d_{\bar q}$, and $d_g$ are taken to be $A$ dependent:
\begin{equation}
\eta = \eta^{(1)} \left ( 1- \frac{1}{A^{\eta^{(2)}}}
       \right ), \ \ \ 
\eta = d_v, \ a_{\bar q}, \ d_{\bar q} , \ d_{g} .
\label{eqn:more-a}
\end{equation}
Because the extreme values $x_{0i}^+$ and $x_{0i}^-$ are almost
independent of $A$ according to the $F_2$ data and also from
the previous analyses \cite{hkm01,hkn04}, they are assumed to be
independent of $A$.
In addition, the parameters $d_{\bar q}^{(1)}$ and $x_{0g}^-$ are
fixed in the antiquark and gluon distributions as follows.
Because the gluon distributions cannot be well determined from
the present data, the parameter $c_g$ is taken $c_g=0$ as assumed
in the previous analysis \cite{hkn04}.
It means that $x_{0g}^-$ and $x_{0g}^+$ are related by
$x_{0g}^- =-  x_{0g}^+$ from Eq. (\ref{eqn:bici}).
There are still six parameters for the antiquark
distributions and they should be too many in comparison with
four parameters for the valence-quark distributions. We decided to
fix the parameter $d_{\bar q}^{(1)}$, which is sensitive to
the gluon shadowing ratio to the antiquark one because of 
the momentum conservation. We found that the gluon shadowing cannot
be well determined even in the NLO analysis; therefore, 
the value of $d_{\bar q}^{(1)}$ is taken so that the gluon shadowing
is similar to the antiquark shadowing.

After all, the following twelve parameters are used for expressing
the nuclear modifications:
\begin{align}
\text{valence quark:}  & \ \ x_{0v}^+,  x_{0v}^-, d_v^{(1)},  d_v^{(2)},
\nonumber \\
\text{antiquark:}      & \ \ x_{0\bar q}^+,  x_{0\bar q}^-, 
                             a_{\bar q}^{(1)},  a_{\bar q}^{(2)},
                             d_{\bar q}^{(2)},  \\
\text{gluon:}          & \ \ x_{0g}^+,  d_g^{(1)},  d_g^{(2)}. 
\nonumber
\label{eqn:paramet}
\end{align}
These parameters are determined by the following global analysis.

\subsection{\label{data} Experimental data}

Most of used experimental data are explained in Ref. \cite{hkn04}.
First, the data for $F_2^A/F_2^D$ are from European Muon Collaboration (EMC)
\cite{emc88, emc90, emc93}, the SLAC (Stanford Linear Accelerator Center)-E49, 
E87, E139, and E140 collaborations \cite{slac83,slac83B,slac88,slac94}, 
the Bologna-CERN-Dubna-Munich-Saclay (BCDMS) collaboration
\cite{bcdms85, bcdms87}, the New Muon Collaboration (NMC) \cite{nmc95},
the Fermilab (Fermi National Accelerator Laboratory)-E665 collaboration
\cite{e665-92,e665-95}, and the HERMES \cite{hermes03}.
Second, the ratios $F_2^{A_1}/F_2^{A_2}$ ($A_2 \ne D$) are from
the NMC \cite{nmc95,nmc96,nmc96snc}.
Third, the Drell-Yan data are from the Fermilab-E772 \cite{e772-90}
and E866/NuSea \cite{e866-99} collaborations.

Additional data to the HKN04 (Hirai, Kumano, Nagai in 2004) analysis 
are the ones for the deuteron-proton ratio $F_2^D/F_2^p$.
These deuteron data are added because precise nuclear modifications
are needed for the deuteron which is, for example,
used in heavy-ion experiments at the Relativistic Heavy Ion Collider
(RHIC) \cite{rhic-heavy-ion}.
The $F_2^D/F_2^p$ data are taken from the measurements by 
the EMC \cite{emc87-d-p}, the BCDMS \cite{bcdms90-d-p},
the Fermilab-E665 \cite{e665-95-d-p}, and the NMC \cite{nmc97-d-p}. 
These data are used for extracting information on the flavor
asymmetric antiquark distributions $(\bar u \ne \bar d)$
in the nucleon \cite{flavor3}. Therefore, the flavor asymmetric
antiquark distributions in Eq. (\ref{eqn:wpart}) are essential
for a successful fit and for extracting information on
modifications in the deuteron.

Because the DGLAP (Dokshitzer-Gribov-Lipatov-Altarelli-Parisi)
evolution can be applied only in the perturbative
QCD region, the data with small $Q^2$ values cannot be used
in the analysis. However, the data in a relatively
small-$Q^2$ region ($Q^2$=1$-$3 GeV$^2$) are valuable for determining
antiquark distributions at small $x$ ($x=0.001-0.01$). As a compromise
of these conflicting conditions, only the data with $Q^2 \ge 1$ GeV$^2$
are used in the analysis. However, one should note that the data
in the range, $Q^2$=1$-$3 GeV$^2$, may contain significant
contributions of higher-twist effects which are not considered
in our leading-twist analysis.

\subsection{\label{chi2analysis} $\chi^2$ analysis}

The parameters are determined by fitting experimental data for
the ratios of the structure functions $F_2^A$ and Drell-Yan cross sections.
The total $\chi^2$
\begin{equation}
\chi^2 = \sum_j \frac{(R_{j}^{\rm data}-R_{j}^{\rm theo})^2}
                     {(\sigma_j^{\rm data})^2},
\label{eqn:chi2}
\end{equation}
is minimized to obtain the optimum parameters.
The ratio $R_j^{\rm data}$ indicates experimental data for
$F_2^A/F_2^{A'}$ and $\sigma_{DY}^{pA}/\sigma_{DY}^{pA'}$,
and $R_j^{\rm theo}$ is a theoretical ratio calculated by the parametrized
NPDFs. The initial scale $Q_0^2$ is taken $Q_0^2$=1 GeV$^2$, and
the distributions in Eq. (\ref{eqn:wpart}) are evolved to experimental
$Q^2$ points to calculate the $\chi^2$ by the DGLAP evolution equations
\cite{dglap-evol}. All the calculations are done in the LO or NLO, and
the modified minimal subtraction ($\overline {\rm MS}$) scheme is used
in the NLO analysis.

The structure function $F_2^A(x,Q^2)$ is expressed in terms of the NPDFs
and coefficient functions:
\begin{align}
F_2^A(x,Q^2) = \sum\limits_{i=1}^{n_f} e_{i}^2  x \, &
\bigg\{ C_q(x,\alpha_s) \otimes [ q_i^A(x,Q^2) +\bar{q}_{i}^A(x,Q^2) ]
\nonumber \\
     & + C_g(x,\alpha_s) \otimes g^A (x,Q^2)  \bigg\},
\label{eqn:f2}
\end{align}
where $C_{q}(x,\alpha_s)$ and $C_{g}(x,\alpha_s)$ are the coefficient
functions \cite{coeff}, and $e_i$ is a quark charge.
The symbol $\otimes$ denotes the convolution integral:
\begin{equation}
f (x) \otimes g (x) = \int^{1}_{x} \frac{dy}{y}
            f\left(\frac{x}{y} \right) g(y) .
\end{equation}

The proton-nucleus Drell-Yan cross section is given by
the summation of $q\bar q$ annihilation and Compton processes
\cite{dy-cross}:
\begin{equation}
\frac{d\sigma^{pA}}{dQ^2 dx_F} = \frac{d\sigma_{q\bar q}^{pA}}{dQ^2 dx_F}
                           +\frac{d\sigma_{qg}^{pA}}{dQ^2 dx_F} .
\end{equation}
They are expressed in terms of the PDFs and subprocess cross sections:
\begin{align} 
& \! \! \! \! \! \! \! \! 
\frac{d\sigma_{q\bar q}^{pA}}{dQ^2 dx_F} = \frac{4\pi \alpha^2}{9Q^2 s}
\sum_i e_i^2 \int_{x_1}^1 dy_1 \int_{x_2}^1 dy_2 \bigg [ 
\frac{d\hat\sigma_{q\bar q(0)}}{dQ^2 dx_F} 
+ \frac{d\hat\sigma_{q\bar q(1)}}{dQ^2 dx_F} \bigg ]
\nonumber \\
& \ \times [ q_i (y_1,Q^2) \bar q_i^A (y_2,Q^2) 
              + \bar q_i (y_1,Q^2)      q_i^A (y_2,Q^2) ],
\label{eqn:dyqqbar}
\end{align}
\begin{align}
& 
\frac{d\sigma_{qg}^{pA}}{dQ^2 dx_F} = \frac{4\pi \alpha^2}{9Q^2 s}
\sum_i e_i^2 \int_{x_1}^1 dy_1 \int_{x_2}^1 dy_2 
\nonumber \\
& \ 
\times 
\bigg [ \frac{d\hat\sigma_{gq}}{dQ^2 dx_F}
   g (y_1,Q^2) [ q_i^A (y_2,Q^2) + \bar q_i^A (y_2,Q^2) ]
\nonumber \\
& \ \ \ 
+  \frac{d\hat\sigma_{qg}}{dQ^2 dx_F}
   [ q_i (y_1,Q^2) + \bar q_i (y_1,Q^2) ] g^A (y_2,Q^2)  \bigg ] .
\label{eqn:dyqg}
\end{align}
The cross sections $d\hat\sigma_{q\bar q(0)}$ and $d\hat\sigma_{q\bar q(1)}$
indicate subprocess cross sections for $q\bar q$ annihilation processes
in the LO and NLO, respectively. The $d\sigma_{qg}$ indicates the cross
section for $qg$ and $\bar q g$ processes. The NLO expressions of these
cross sections are, for example, found in Ref. \cite{dy-cross}.
Effects of possible parton-energy loss in the Drell-Yan process
\cite{duan-energy-loss} are neglected in this analysis.

\begin{table*}[t]
\caption{\label{table:parameters}
         Parameters obtained by the LO and NLO analyses.
         The parameters $a_{u_v}$, $a_{d_v}$, and $a_{g}$ are fixed by
         the three conservation conditions in Eq. (\ref{eqn:3conserv}).
         Since they depend on nuclear species,
         they are explained separately in Appendix \ref{appen-a}.}
\begin{ruledtabular}
\begin{tabular}{ccccccc} 
Distribution   & $x_0^+$                & $x_0^-$
               & $d^{(1)}$              & $d^{(2)}$
               & $a^{(1)}$              & $a^{(2)}$              \\
\hline
(LO)             & & & & & & \\
$u_v^A$, $d_v^A$ & 0.675    $\pm$ 0.005  & 0.208    $\pm$ 0.004
                 & 5.089    $\pm$ 0.315  & 0.876    $\pm$ 0.143
                 & (Appendix)            & (Appendix)            \\
$\bar q^A$       & 0.120    $\pm$ 0.007  & 0.387    $\pm$ 0.027
                 & 60 (fixed)            & 0.260    $\pm$ 0.043  
                 & $-$0.394 $\pm$ 0.027  & 0.318    $\pm$ 0.038  \\
$g^A$            & 8.35     $\pm$ 9.09   & $-x_0^+$ (fixed)  
                 & $-$799   $\pm$ 1561
                 & (0.194 $\pm$ 0.341) $\times 10^{-5}$
                 & (Appendix)             & (Appendix)           \\
\hline
(NLO)            & & & & & & \\
$u_v^A$, $d_v^A$ & 0.629    $\pm$ 0.004  & 0.178    $\pm$ 0.004
                 & 6.157    $\pm$ 0.254  & 1.063    $\pm$ 0.143
                 & (Appendix)            & (Appendix)            \\
$\bar q^A$       & 0.239    $\pm$ 0.009  & 0.119    $\pm$ 0.004
                 & 150 (fixed)           & 0.212    $\pm$ 0.019  
                 & $-$0.540 $\pm$ 0.024  & 0.295    $\pm$ 0.024  \\
$g^A$            & 7.24     $\pm$ 3.97   & $-x_0^+$ (fixed)  
                 & $-$26.5  $\pm$ 27.8 
                 & (0.109 $\pm$ 0.114) $\times 10^{-3}$
                 & (Appendix)             & (Appendix)           \\
\end{tabular}
\end{ruledtabular}
\end{table*}

Using these expressions for the structure functions $F_2^A$ and
Drell-Yan cross sections, we calculate the theoretical ratios
$R_j^{\rm theo}$ in Eq. (\ref{eqn:chi2}). The total $\chi^2$ is minimized
by the CERN program library {\tt MINUIT}. From this analysis,
an error matrix which is the inverse of a Hessian matrix, is obtained.
NPDF uncertainties are estimated by using the Hessian matrix as 
\begin{equation}
\! \! \! \!  
        [\delta f^A(x)]^2=\Delta \chi^2 \sum_{i,j}
          \left( \frac{\partial f^A(x,\xi)}{\partial \xi_i}  
                \right) _{\xi=\hat{\xi}}
          H_{ij}^{-1}
          \left( \frac{\partial f^A(x,\xi)}{\partial \xi_j} 
                \right) _{\xi=\hat{\xi}}  ,
        \label{eq:dnpdf}
\end{equation}
where $H_{ij}$ is the Hessian matrix, $\xi_i$ is a parameter, 
and $\hat\xi$ indicates the optimum parameter set.
The $\Delta \chi^2 $ value determines the confidence region,
and it is calculated so that the confidence level $P$
becomes the one-$\sigma$-error range ($P=0.6826$) for a given
number of parameters ($N$) by assuming the normal distribution
in the multiparameter space.
In the analysis with the twelve parameters, it is $\Delta \chi^2 =13.7$.
This Hessian method has been used for estimating polarized PDFs
and fragmentation functions \cite{aac0306,hkns07} as well as
nuclear PDFs in our previous version \cite{hkn04}.
The details of the uncertainty analysis are discussed in
Refs. \cite{aac0306,hkn04} as well as in the nucleonic PDF articles 
\cite{unpol-error}.

\section{\label{results} Results}

\subsection{Comparison with data}
\label{x-dependence} 

Determined parameters are listed in Table \ref{table:parameters} for both
LO and NLO. Three parameters are fixed by the constraints from baryon-number,
charge, and momentum conservations in Eq. (\ref{eqn:3conserv}),
and we chose $a_{u_v}$, $a_{d_v}$, and $a_g$ for these parameters.
The values of the obtained parameters
and their errors are similar in the LO and NLO. However, the errors indicate
that there are slight NLO improvements in comparison with the LO results.

\begin{table}[h]
\caption{Each $\chi^2$ contribution.}
\label{tab:chi2}
\begin{ruledtabular}
\begin{tabular*}{\hsize}
{c@{\extracolsep{0ptplus1fil}}c@{\extracolsep{0ptplus1fil}}c
@{\extracolsep{0ptplus1fil}}c@{\extracolsep{0ptplus1fil}}c}
Nucleus  & Reference & \# of data & $\chi^2$      & $\chi^2$  \\
         &           &            &  (LO)         &  (NLO)    \\
\colrule
D/p      & \cite{emc87-d-p,bcdms90-d-p,e665-95-d-p,nmc97-d-p}  
         &      290    &       375.5    &       322.5         \\
\colrule
$^4$He/D & \cite{slac94,nmc95}
         &   \   35    &  \     60.9    &  \     51.8         \\
Li/D     & \cite{nmc95} 
         &   \   17    &  \     36.9    &  \     36.4         \\ 
Be/D     & \cite{slac94}
         &   \   17    &  \     39.3    &  \     53.0         \\
C/D      & \cite{emc88,emc90,slac94,nmc95,e665-95} 
         &   \   43    &       105.8    &  \     78.7         \\
N/D      & \cite{bcdms85,hermes03}
         &   \  162    &       136.3    &       121.7         \\
Al/D     & \cite{slac83B,slac94}
         &   \   35    &  \     45.5    &  \     44.9         \\
Ca/D     & \cite{emc90,nmc95,slac94,e665-95}
         &   \   33    &  \     43.3    &  \     34.1         \\
Fe/D     & \cite{slac83,slac88,slac94,bcdms87}
         &   \   57    &       108.0    &  \     97.4         \\
Cu/D     & \cite{emc93} 
         &   \   19    &  \     12.1    &  \     13.2         \\
Kr/D     & \cite{hermes03}
         &      144    &       115.1    &       115.9         \\
Ag/D     & \cite{slac94}
         &   \ \  7    &  \     12.5    &  \ \    9.1         \\
Sn/D     & \cite{emc88} 
         &   \ \  8    &  \     13.3    &  \     14.1         \\
Xe/D     & \cite{e665-92} 
         &   \ \  5    &  \ \    2.2    &  \ \    2.3         \\
Au/D     & \cite{slac88,slac94}
         &   \   19    &  \     55.6    &  \     32.3         \\
Pb/D     & \cite{e665-95}
         &   \ \  5    &  \ \    5.7    &  \ \    4.5         \\
\colrule
$F_2^A/F_2^D$ total
      &  &      606    &       792.4    &       709.3         \\
\colrule
Be/C     & \cite{nmc96}
         &   \   15    &  \     12.6    &  \     11.9         \\
Al/C     & \cite{nmc96} 
         &   \   15    &  \ \    5.0    &  \ \    5.1         \\
Ca/C     & \cite{nmc95,nmc96}
         &   \   39    &  \     29.9    &  \     29.1         \\
Fe/C     & \cite{nmc96}
         &   \   15    &  \ \    8.0    &  \ \    8.3         \\
Sn/C     & \cite{nmc96snc}
         &      146    &       204.0    &       172.1         \\
Pb/C     & \cite{nmc96}
         &   \   15    &  \     15.7    &  \     12.2         \\
C/Li     & \cite{nmc95}
         &   \   24    &  \     67.4    &  \     64.9         \\
Ca/Li    & \cite{nmc95}
         &   \   24    &  \     69.0    &  \     65.3         \\
\colrule
$F_2^{A_1}/F_2^{A_2}$ total 
      &  &      293    &       411.6    &       369.0         \\
\colrule
C/D      & \cite{e772-90} 
         &   \ \  9    &  \ \    9.3    &  \ \    8.1         \\
Ca/D     & \cite{e772-90} 
         &   \ \  9    &  \ \    5.8    &  \     13.8         \\
Fe/D     & \cite{e772-90} 
         &   \ \  9    &  \     12.6    &  \     17.9         \\
W/D      & \cite{e772-90} 
         &   \ \  9    &  \     27.8    &  \     29.6         \\
Fe/Be    & \cite{e866-99}
         &   \ \  8    &  \ \    3.3    &  \ \    3.6         \\
W/Be     & \cite{e866-99}
         &   \ \  8    &  \     14.9    &  \     12.1         \\
\colrule
Drell-Yan total  
      &  &   \   52    &  \     73.8    &  \     85.1         \\
\colrule\colrule
Total &  &    1241    &      1653.3 \  &      1485.9 \       \\
($\chi^2$/d.o.f.)  & & &       (1.35)   &      (1.21)         \\
\end{tabular*}
\end{ruledtabular}
\end{table}

Each $\chi^2$ contribution is listed in Table \ref{tab:chi2}.
The values suggest that medium and large nuclei should be well explained
by the current LO and NLO parametrizations. However, small nuclei
are not so well reproduced. The LO fit ($\chi^2$/d.o.f.=1.35) is better
than the previous analysis with $\chi^2$/d.o.f.=1.58 \cite{hkn04},
which is partly due to the introduction of new parameters
for the $A$ dependence in Eq. (\ref{eqn:more-a}).
If the $\chi^2$ values of the LO analysis are compared with
the ones in Table I\hspace{-.1em}I\hspace{-.1em}I of Ref. \cite{hkn04},
we find that much improvements are obtained for the nuclear ratios,
$F_2^{Li}/F_2^D$, $F_2^C/F_2^D$, 
$F_2^{Be}/F_2^C$, $F_2^{Ca}/F_2^C$, $F_2^{Fe}/F_2^C$, $F_2^{Sn}/F_2^C$,
$F_2^{Pb}/F_2^C$, $F_2^{Ca}/F_2^{Li}$, 
$\sigma_{DY}^{pCa}/\sigma_{DY}^{pD}$, 
$\sigma_{DY}^{pFe}/\sigma_{DY}^{pBe}$, and
$\sigma_{DY}^{pW} /\sigma_{DY}^{pBe}$,
whereas the fit becomes worse for
$\sigma_{DY}^{pFe}/\sigma_{DY}^{pD}$ and
$\sigma_{DY}^{pW}/\sigma_{DY}^{pD}$.
The $\chi^2$/d.o.f. is further reduced in the current NLO analysis.

According to Table \ref{tab:chi2}, the NLO results with $\chi^2$/d.o.f.=1.21
reproduce the data better than the LO ones with $\chi^2$/d.o.f.=1.35,
especially in the following data sets:
$F_2^{D}/F_2^p$, $F_2^{^4 He}/F_2^D$, $F_2^{C}/F_2^D$,
$F_2^{Ca}/F_2^D$, $F_2^{Ag}/F_2^D$, $F_2^{Au}/F_2^D$, $F_2^{Pb}/F_2^D$, 
$F_2^{Sn}/F_2^C$, $F_2^{Pb}/F_2^C$, and
$\sigma_{DY}^{pW}/\sigma_{DY}^{pBe}$,
whereas it becomes worse in $F_2^{Be}/F_2^D$,
$\sigma_{DY}^{pCa}/\sigma_{DY}^{pD}$, and
$\sigma_{DY}^{pFe}/\sigma_{DY}^{pD}$. 
The deuteron-proton data $F_2^D/F_2^p$ are added in this analysis
to the data set of the previous version \cite{hkn04}, and they
should provide a valuable constraint on PDF modifications in the deuteron.
Because the $F_2^D/F_2^p$ data are sensitive to $\bar u/\bar d$ asymmetry
\cite{flavor3}, flavor asymmetric antiquark distributions should be used
in our analysis. If the flavor symmetric distributions are used
as initial ones, the fit produces a significantly larger $\chi^2$. 

\begin{figure}[t]
        \includegraphics*[width=42.3mm]{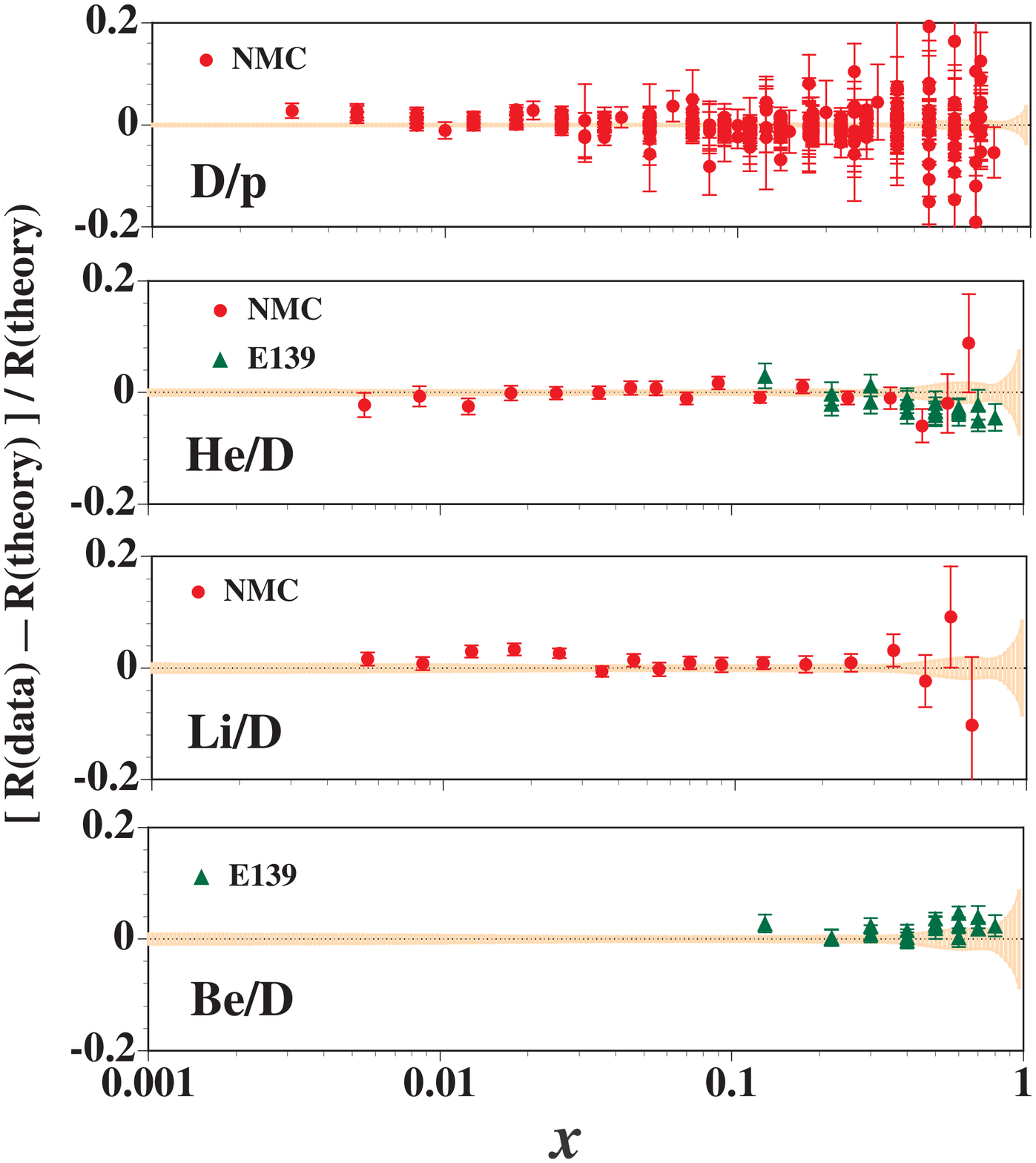} \hspace{0.5mm}
        \includegraphics*[width=39.0mm]{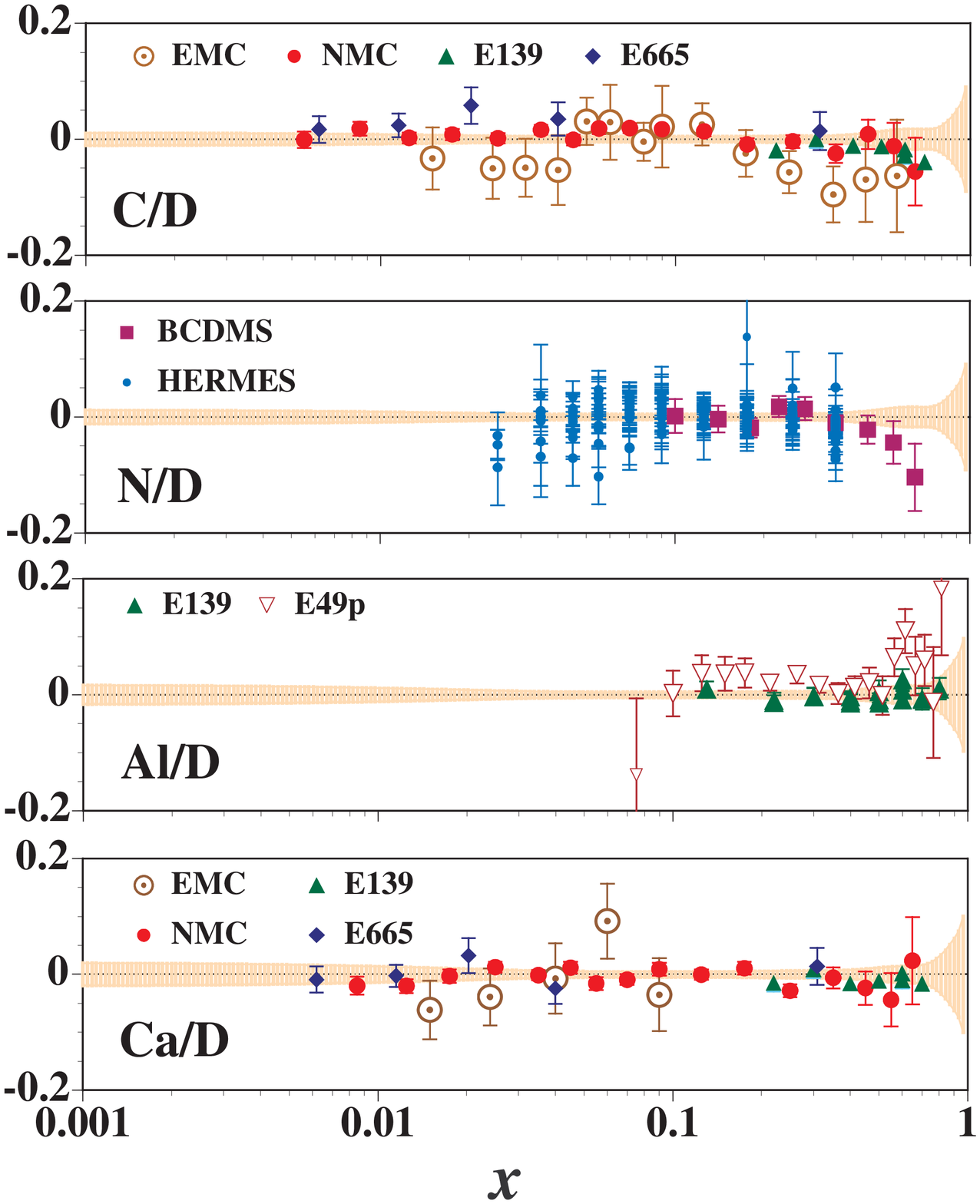} \\
\vspace{0.5cm}
        \includegraphics*[width=42.3mm]{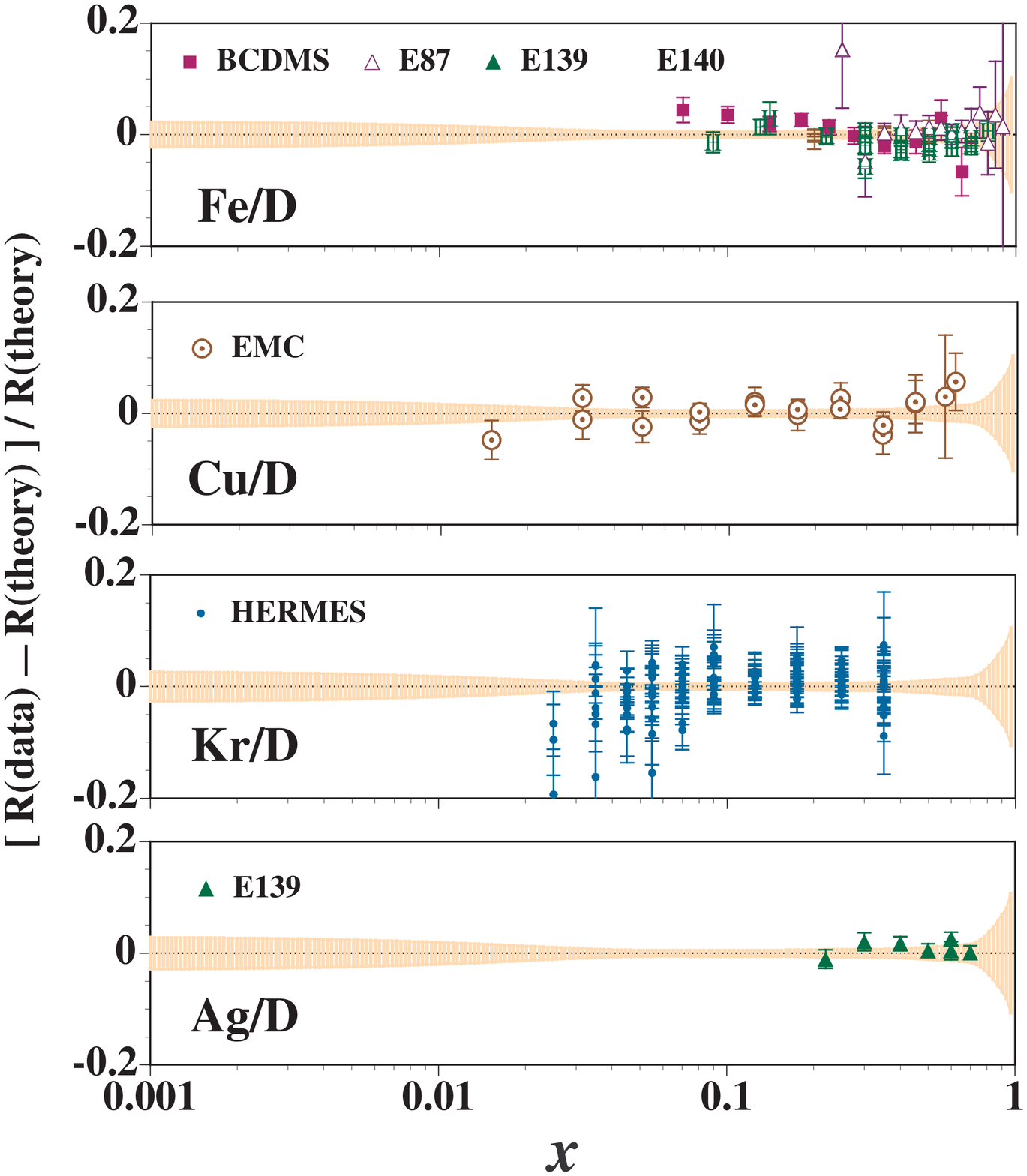} \hspace{0.5mm}
        \includegraphics*[width=39.0mm]{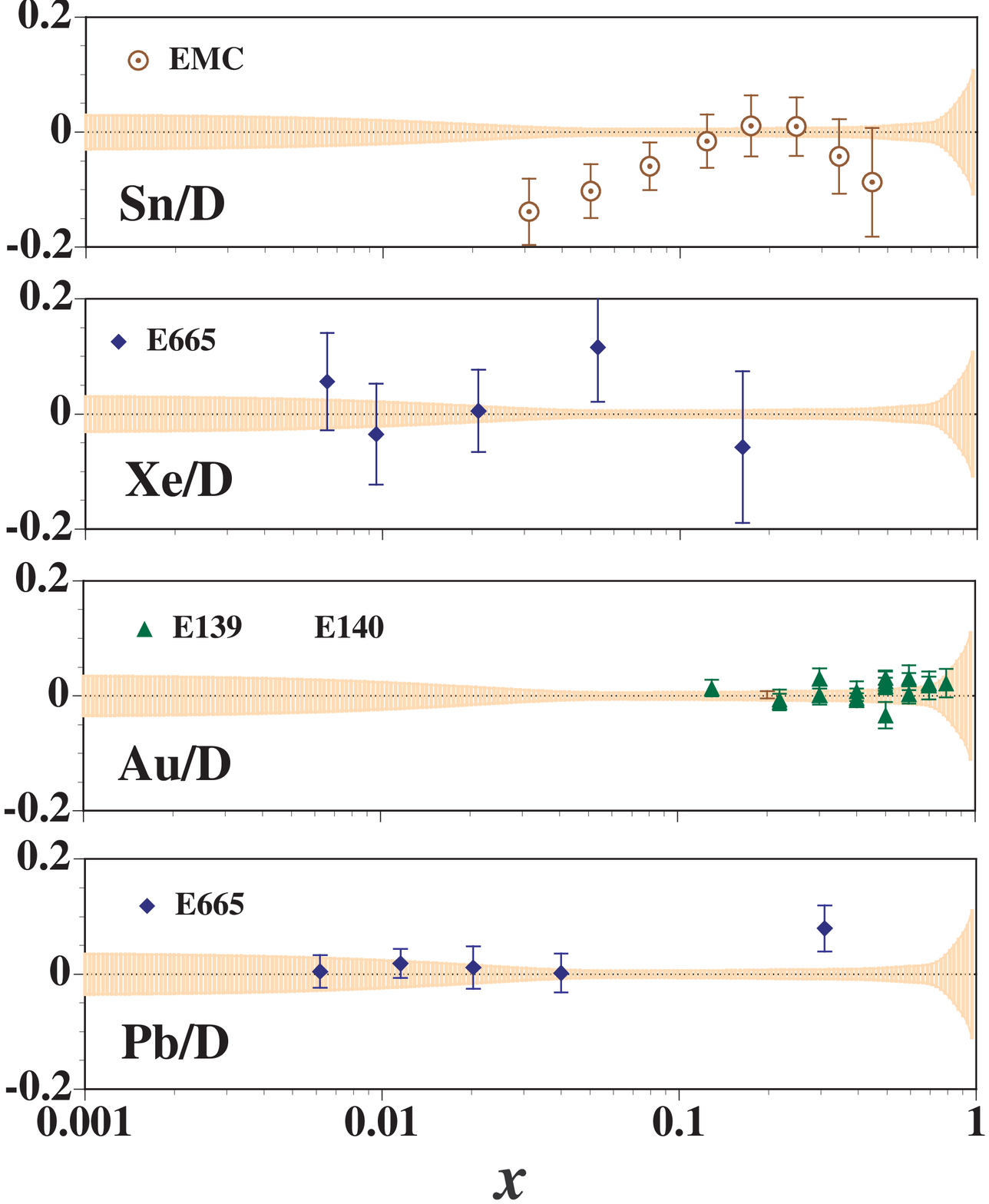} 
       \vspace{-0.25cm}
\caption{\label{fig:RD}
(Color online) Comparison with experimental ratios $R=F_2^A/F_2^D$
and $F_2^D/F_2^p$. The rational differences between experimental
and theoretical values [$(R^{\rm exp}-R^{\rm theo})/R^{\rm theo}$]
are shown.
The NLO parametrization is used for the theoretical calculations
at the $Q^2$ points of the experimental data. Theoretical uncertainties
in the NLO are shown at $Q^2$=10 GeV$^2$ by the shaded areas.}
\end{figure}

\begin{figure}[t]
        \includegraphics*[width=42.3mm]{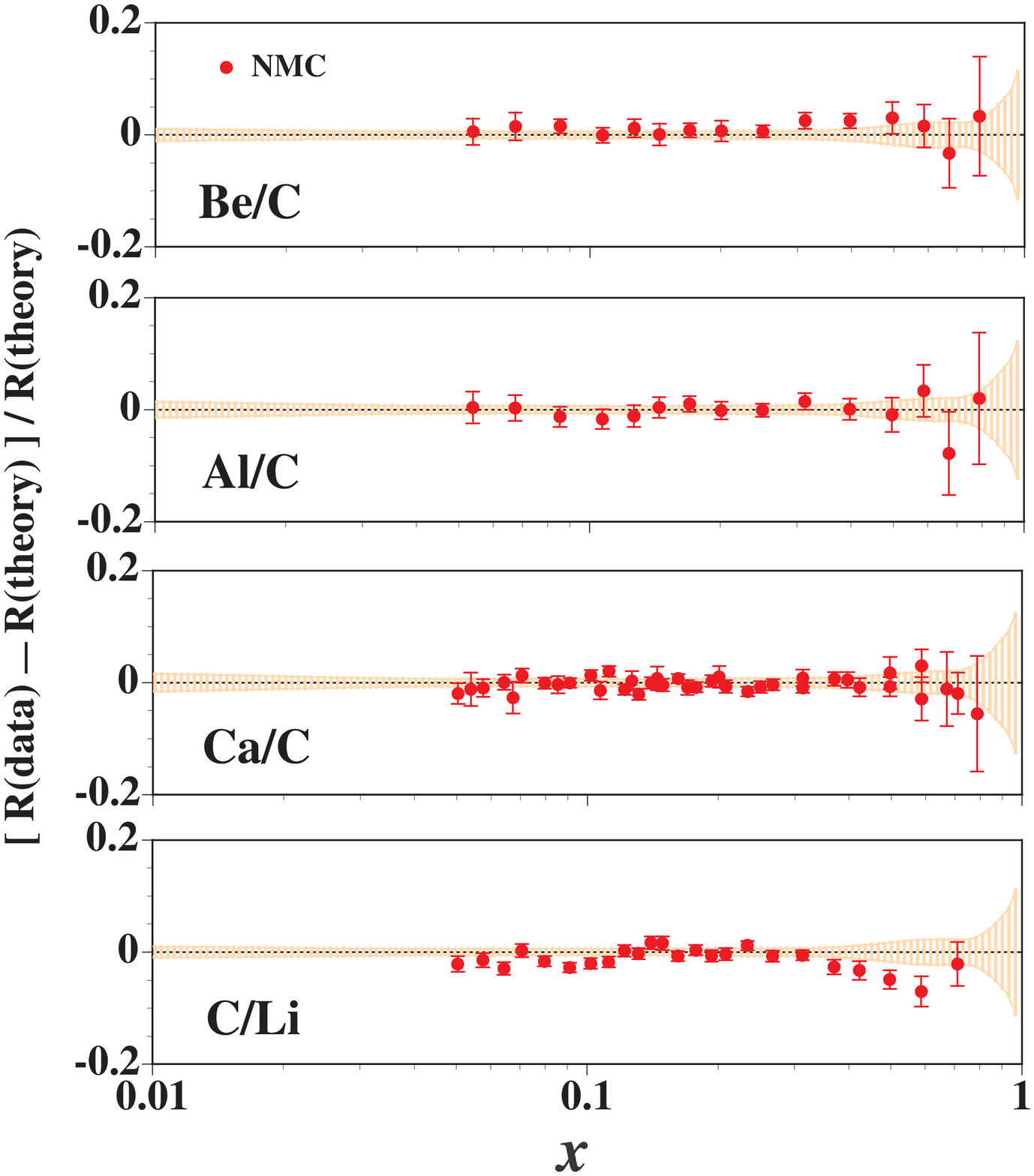} \hspace{0.5mm}
        \includegraphics*[width=39.0mm]{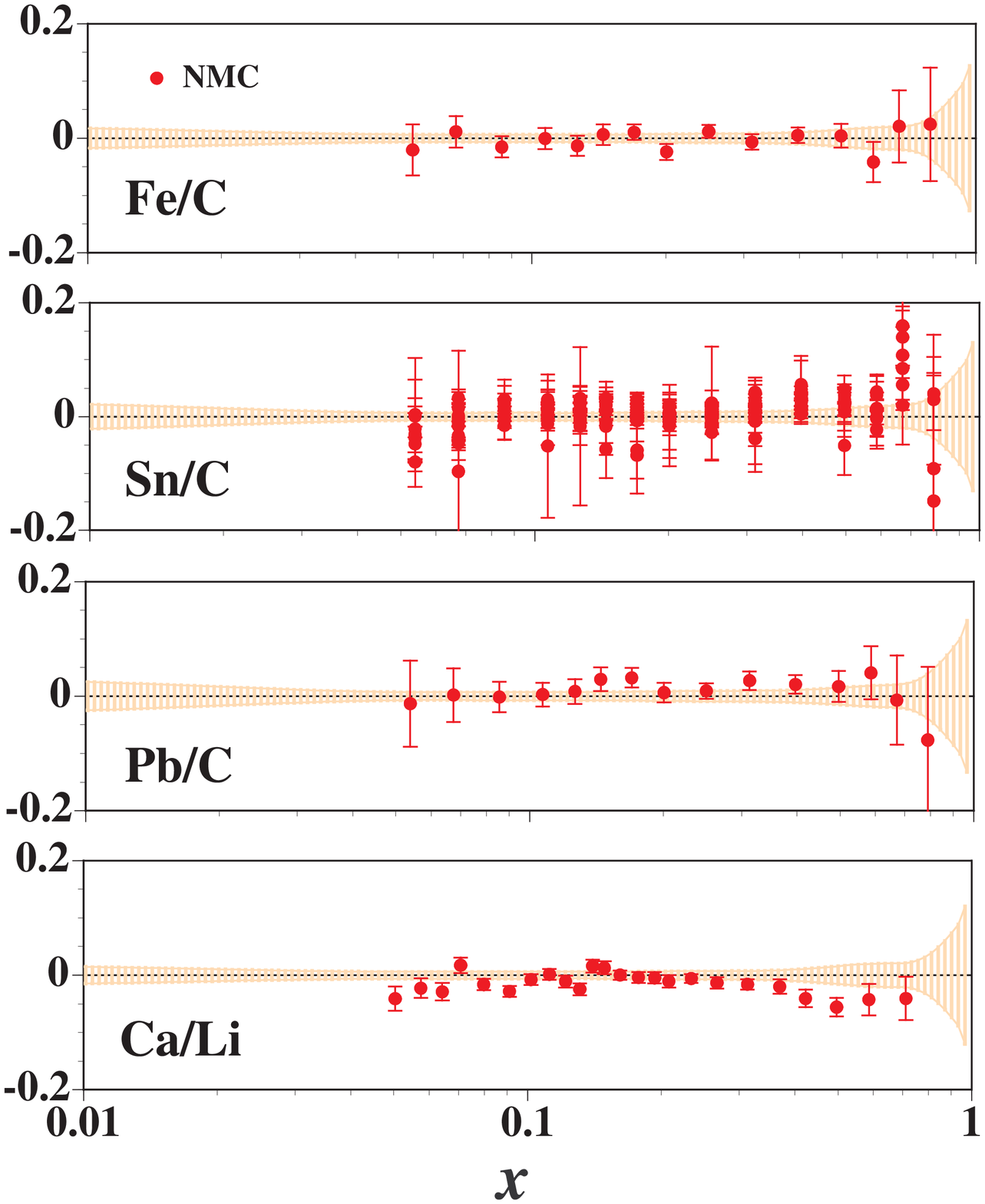} \\
       \vspace{-0.25cm}
\caption{\label{fig:RA}
(Color online) Comparison with experimental data of $R=F_2^A/F_2^{C,Li}$. 
The ratios $(R^{\rm exp}-R^{\rm theo})/R^{\rm theo}$ are shown.
The theoretical ratios and their uncertainties are calculated in the NLO.
The notations are the same as Fig. \ref{fig:RD}.}
\end{figure}

\begin{figure}[t]
        \includegraphics*[width=42.3mm]{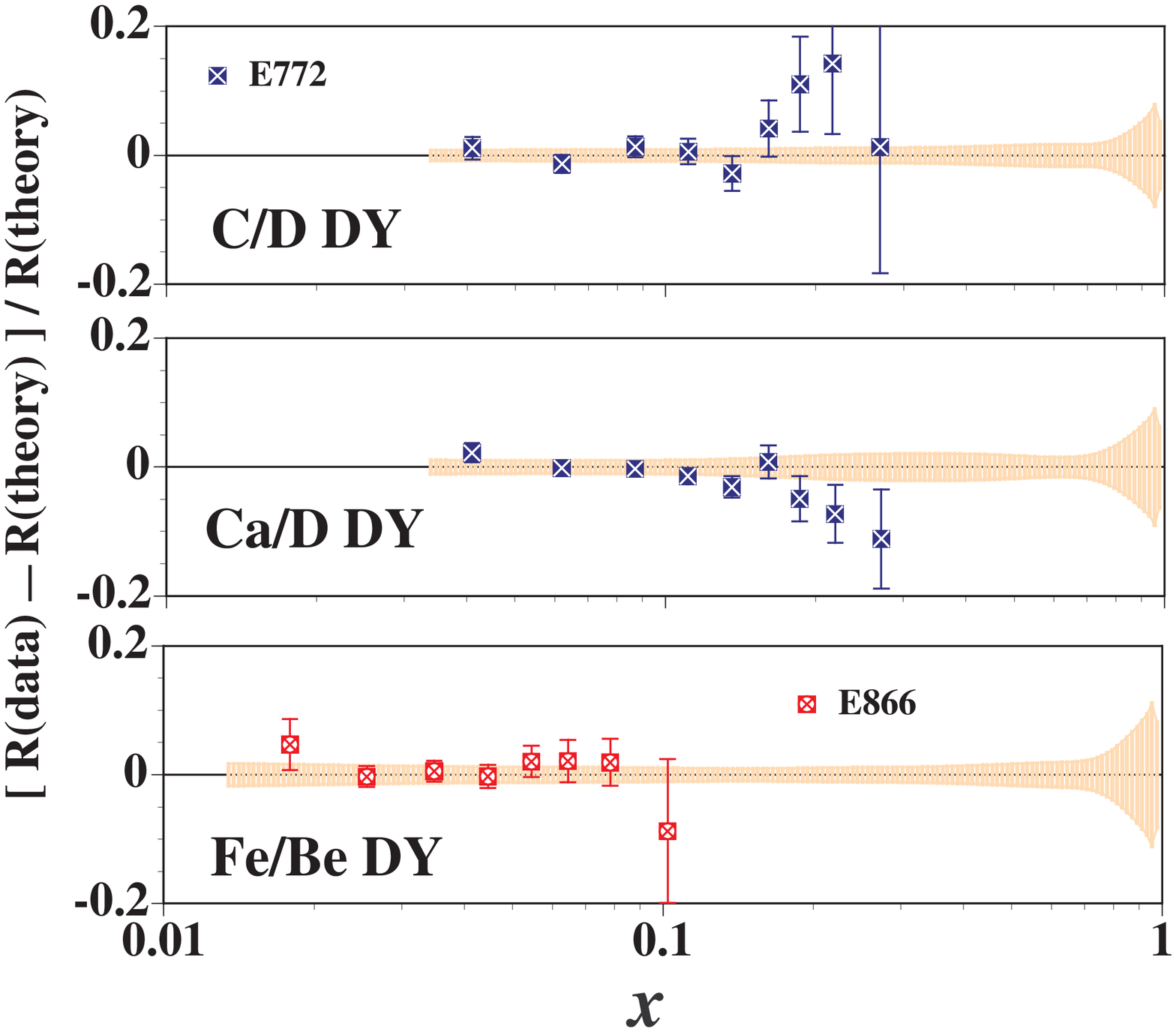} \hspace{0.5mm}
        \includegraphics*[width=39.0mm]{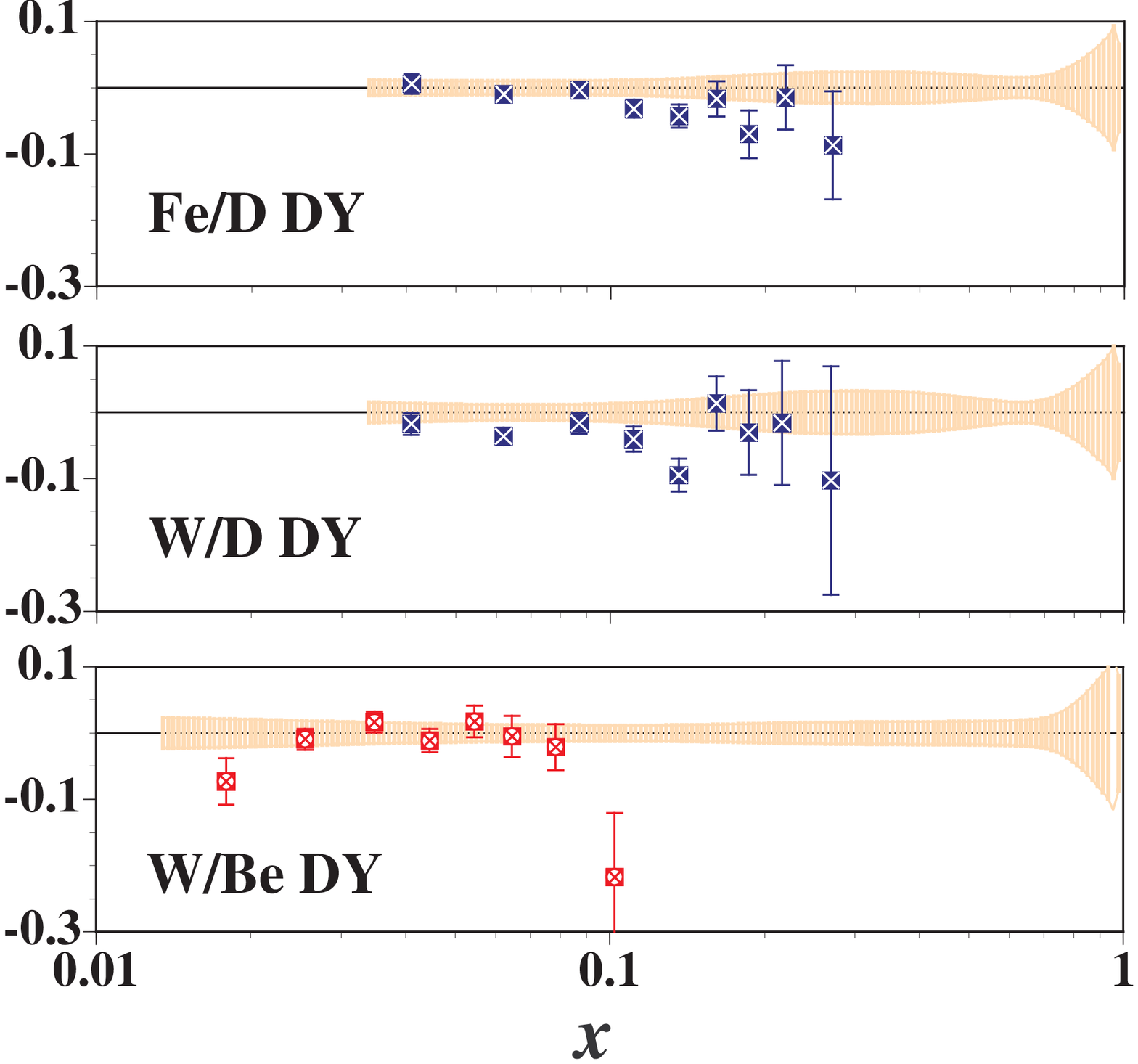} \\
       \vspace{-0.25cm}
\caption{\label{fig:DY}
(Color online) Comparison with Drell-Yan data of
$R=\sigma_{DY}^{pA}/\sigma_{DY}^{pA'}$.
The ratios $(R^{\rm exp}-R^{\rm theo})/R^{\rm theo}$ are shown.
The theoretical ratios and their uncertainties are calculated in the NLO.
The theoretical ratios are calculated at the $Q^2$ points of
the experimental data. The uncertainties are estimated at $Q^2$=20 and 50
GeV$^2$ for the the $\sigma_{DY}^{pA}/\sigma_{DY}^{pBe}$ type and 
$\sigma_{DY}^{pA}/\sigma_{DY}^{pD}$ one, respectively.}
\end{figure}

The fit results of the NLO are compared with the used data in
Figs. \ref{fig:RD}, \ref{fig:RA}, and \ref{fig:DY} for
the ratios $F_2^A/F_2^D$, $F_2^{A_1}/F_2^{A_2}$, and
$\sigma_{DY}^{pA_1}/\sigma_{DY}^{pA_2}$, respectively.
The rational differences between experimental and theoretical
values $(R^{\rm exp}-R^{\rm theo})/R^{\rm theo}$, where $R$ is 
$R=F_2^A/F_2^D$, $F_2^{A_1}/F_2^{A_2}$, or 
$\sigma_{DY}^{pA_1}/\sigma_{DY}^{pA_2}$,
are shown. For the theoretical values, the NLO results are used and
they are calculated at the experimental $Q^2$ points. The uncertainty
bands are also shown in the NLO, and they are calculated at
$Q^2$=10 GeV$^2$ for the structure function $F_2$ and at $Q^2$=20 or
50 GeV$^2$ for the Drell-Yan processes. 
The scale $Q^2$=10 GeV$^2$ is taken because the average of all the $F_2$
data is of the order of this value. The scale is $Q^2$=50 GeV$^2$
for the Drell-Yan ratios of the $\sigma_{DY}^{pA}/\sigma_{DY}^{pD}$ type,
and the lower scale 20 GeV$^2$ is taken for the ratio of 
the $\sigma_{DY}^{pA}/\sigma_{DY}^{pBe}$ type
because experimental $Q^2$ values are smaller.

These figures indicate that the overall fit is successful
in explaining the used data. We notice that the $\chi^2$ values,
53.0, 64.9, and 29.6 in the NLO, are especially large for
$F_2^{Be}/F_2^{D}$, $F_2^{C}/F_2^{Li}$, and $\sigma_{DY}^{pW}/\sigma_{DY}^{pD}$
in comparison with the numbers of their data, 17, 24, and 9,
according to Table \ref{tab:chi2}. These large $\chi^2$ values
come from deviations from accurate E139, NMC, and E772 data; however,
such deviations are not very significant in Figs. \ref{fig:RD},
\ref{fig:RA}, and \ref{fig:DY}. There are general tendencies that
medium- and large-size nuclei are well explained by our parametrization,
whereas there are slight deviations for small nuclei.
Because any systematic deviations are not found from the experimental data,
our analyses should be successful in determining the optimum nuclear PDFs.

\begin{figure}[t]
        \includegraphics*[width=42.5mm]{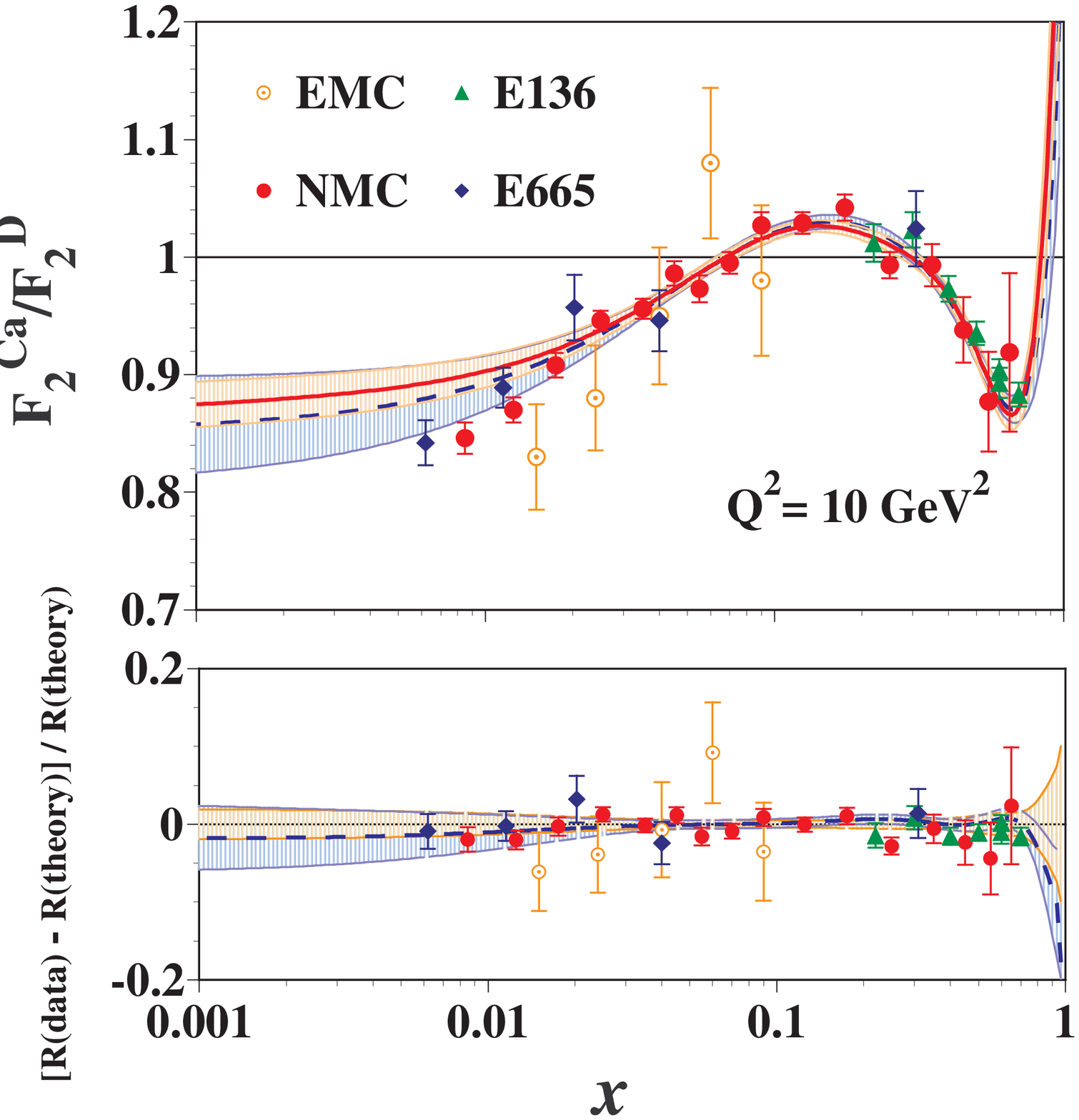} \hspace{0.5mm}
        \includegraphics*[width=41.0mm]{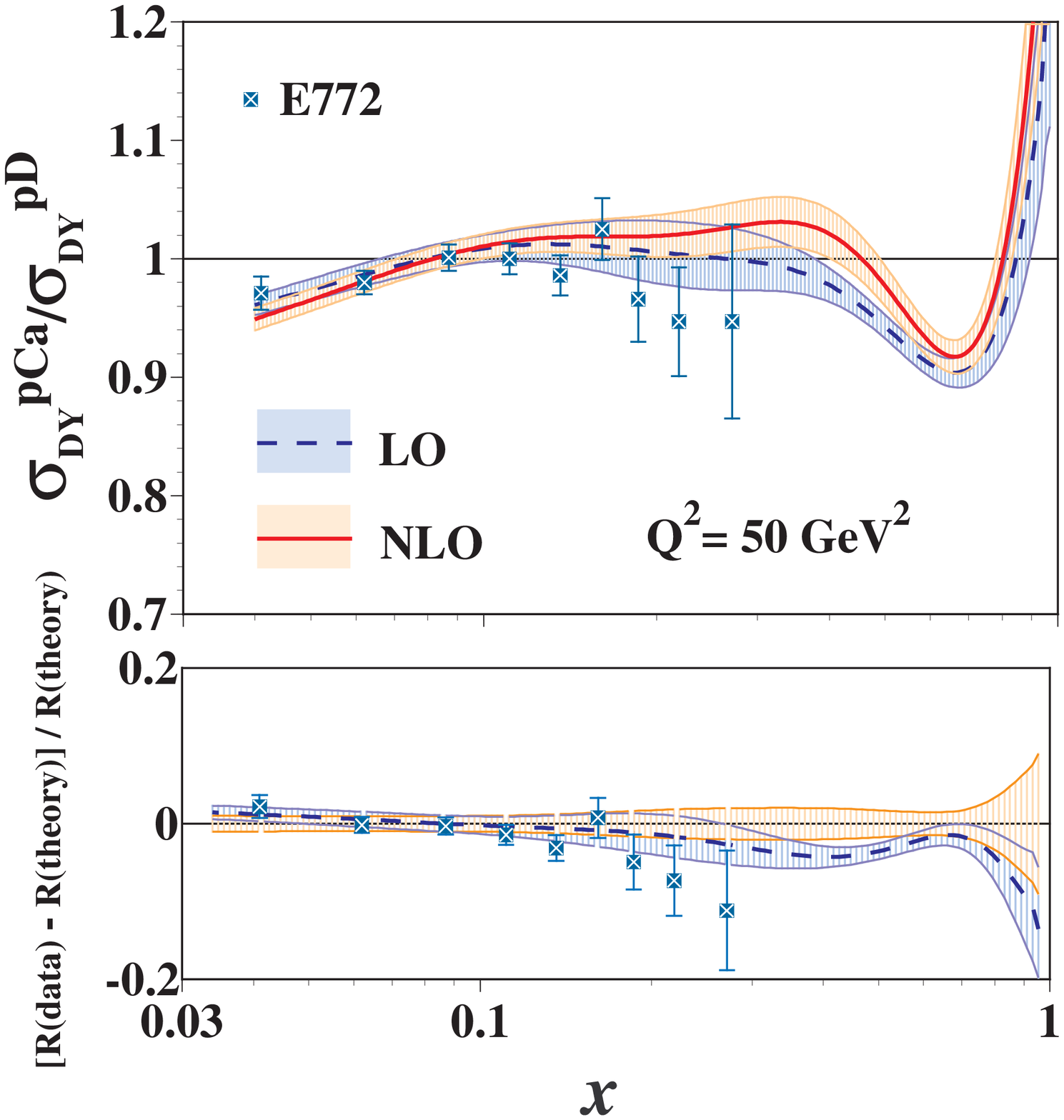} \\
        \vspace{-0.3cm}
\caption{(Color online) Theoretical results are compared with
         the data of the $F_2$ ratio $F_2^{Ca}/F_2^{D}$
         and the Drell-Yan ratio $\sigma_{DY}^{pCa}/\sigma_{DY}^{pD}$.
         In the upper figures,
         the theoretical curves and uncertainties are calculated 
         at $Q^2$=10 GeV$^2$ for the $F_2$ ratio and 
         at $Q^2$=50 GeV$^2$ for the Drell-Yan ratio.
         The dashed and solid curves indicate LO and NLO results, and
         the LO and NLO uncertainties are shown by the dark- and
         light-shaded bands, respectively.
     The lower figures indicate the ratios
     $(R^{\rm exp}-R^{\rm theo})/R^{\rm theo}$ where $R$ indicates
     $F_2^{Ca}/F_2^D$ or $\sigma_{DY}^{pCa}/\sigma_{DY}^{pD}$.
     Here, the theoretical ratios are calculated at the experimental
     $Q^2$ points. For comparison, the LO curves and their
     uncertainties are also shown by 
     $R^{\rm theo}({\rm LO})/R^{\rm theo}({\rm NLO})-1$ and
     $\delta R^{\rm theo}({\rm LO})/R^{\rm theo}({\rm NLO})$.}
\label{fig:f2-dy}
\end{figure}

Next, actual data are compared with the LO and NLO theoretical ratios and
their uncertainties for the calcium nucleus as an example in
Fig. \ref{fig:f2-dy}. 
In the upper figures, the theoretical curves and the uncertainties
are calculated at fixed $Q^2$ points, $Q^2$=10 GeV$^2$ and
50 GeV$^2$ for the $F_2$ and the Drell-Yan, respectively,
whereas the experimental data are taken at various $Q^2$ values.
The rational differences $(R^{\rm exp}-R^{\rm theo})/R^{\rm theo}$
are shown together with the difference between the LO and NLO curves,
$R^{\rm theo}({\rm LO})/R^{\rm theo}({\rm NLO})-1$, in the lower figures.
The comparison suggests that
both LO and NLO parametrizations should be successful in explaining
the $x$ dependence of the calcium data. It is noteworthy that 
the NLO error band of the $F_2$ ratio becomes slightly smaller
in comparison with the LO one at small $x$; however, magnitudes of
both uncertainties are similar in the region, $x>0.02$. 
The NLO improvement is not clearly seen in the Drell-Yan ratio
$\sigma_{DY}^{pCa}/\sigma_{DY}^{pD}$ in the range of $x>0.04$.
There are discrepancies between the theoretical curves and
the $F_2^{Ca}/F_2^{D}$ data at $x<0.01$; however, they are simply
due to $Q^2$ differences. If the theoretical ratios are calculated
at the same experimental $Q^2$ points, they agree as shown in
the $F_2^{Ca}/F_2^{D}$ part of Fig. \ref{fig:RD}.

These LO and NLO results indicate that the available data are taken
in the limited $x$ range without small-$x$ data, and they are not
much sensitive to NLO corrections. This fact leads to a difficulty
in determining nuclear gluon distributions because the gluonic
effects are typical NLO effects through the coefficient functions
and in the $Q^2$ evolution equations. 

\subsection{$Q^2$ dependence}
\label{q2-dependence} 

One of possible methods for determining the gluon distribution
in the nucleon is to investigate $Q^2$ dependence of the structure
function $F_2$ \cite{f2-gluon}. Because $Q^2$ dependent data exist
in the $F_2^{A}/F_2^{A'}$ ratios, it may be possible to find
nuclear modifications of the gluon distribution. 

\begin{figure}[t]
\includegraphics[width=0.41\textwidth]{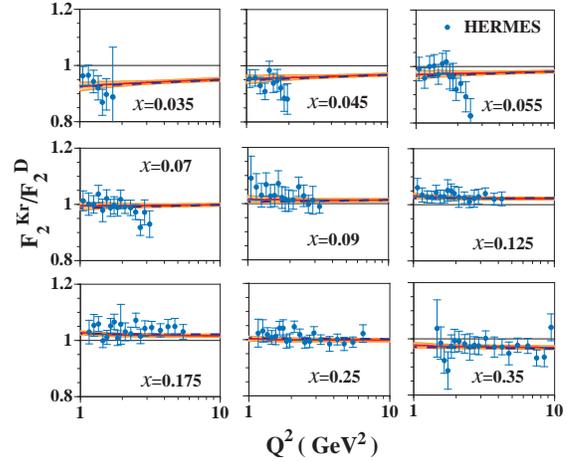}
\vspace{-0.2cm}
\caption{(Color online) Comparison with $Q^2$-dependent data
    of $F_2^{Kr}/F_2^{D}$ by the HERMES collaboration.
    The dashed and solid curves indicate LO and NLO results, and
    the NLO uncertainties are shown by the shaded bands.}
\label{fig:krd-q2}
\end{figure}

\begin{figure}[t]
\vspace{-0.0cm}
\includegraphics[width=0.41\textwidth]{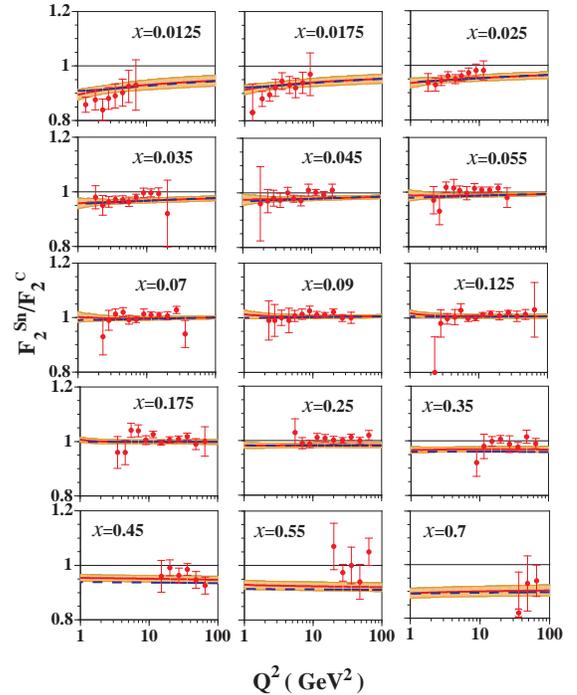}
\vspace{-0.2cm}
\caption{(Color online) Comparison with $Q^2$-dependent data
    of $F_2^{Sn}/F_2^{C}$ by the NMC. The notations
    are the same as Fig. \ref{fig:krd-q2}.}
\label{fig:snc-q2}
\vspace{-0.2cm}
\end{figure}

The LO and NLO parametrization results are compared with
$Q^2$ dependent data for $F_2^{Kr}/F_2^{D}$ and $F_2^{Sn}/F_2^{C}$
measured by the HERMES and NMC collaborations, respectively,
in Figs. \ref{fig:krd-q2} and \ref{fig:snc-q2}. The uncertainties
are shown by the shaded bands in the NLO. Because the LO and NLO
uncertainties are similar except for the small-$x$ region, the LO
ones are not shown in these figures. They are compared later
in this subsection.
The results indicate that overall $Q^2$ dependencies are well
explained by our parametrizations in both LO and NLO. The comparison
suggests that the experimental data are not accurate enough to probe
the details of the $Q^2$ dependence. 
Furthermore, $Q^2$ dependencies of the HERMES and NMC results are different.
The HERMES ratio $F_2^{Kr}/F_2^D$ tends to decrease with increasing $Q^2$ 
at $x$=0.035, 0.045, and 0.055, whereas the NMC ratio $F_2^{Sn}/F_2^C$ 
increases with $Q^2$ at the same $x$ points, although the nuclear species 
are different. This kind of difference together with inaccurate
$Q^2$-dependent measurements makes it difficult to extract precise nuclear
gluon distributions within the leading-twist DGLAP approach.
It is reflected in large uncertainties in the gluon distributions
as it becomes obvious in Sec. \ref{npdfs}.

In our previous versions \cite{hkm01,hkn04}, the experimental
shadowing in $F_2^{Sn}/F_2^C$ is underestimated at small $x$
($0.01<x<0.02$) partly because of an assumption on a simple
$A$ dependence. As shown in Fig. \ref{fig:snc-q2}, the shadowing
is still slightly underestimated at $x=0.0125$; however, the deviations
are not as large as before. If the experimental errors and the NPDF
uncertainties are considered, our parametrization is consistent with
the data.

\begin{figure}[t]
\vspace{-0.0cm}
\includegraphics[width=0.40\textwidth]{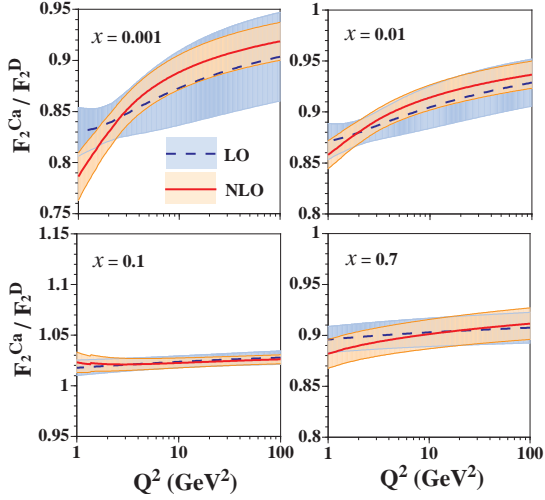}
\vspace{-0.2cm}
\caption{(Color online) $Q^2$ dependence of the ratio $F_2^{Ca}/F_2^{D}$ 
 is compared in the LO and NLO at $x$=0.001, 0.01, 0.01, and 0.7.
 The dashed and solid curves indicate LO and NLO results, and
 LO and NLO uncertainties are shown by the dark- and
 light-shaded bands, respectively.}
\label{fig:cad-q2}
\vspace{-0.2cm}
\end{figure}

The NLO uncertainties are compared with the LO ones in Fig. \ref{fig:cad-q2}
for the ratio $F_2^{Ca}/F_2^{D}$. The LO and NLO ratios and their
uncertainties are shown at $x$=0.001, 0.01, 0.1, and 0.7.
The differences between both uncertainties are conspicuous
at small $x$ (=0.001 and 0.01); however, they are similar
at larger $x$. The LO and NLO slopes are also different at small $x$.
These results indicate that the NLO effects become important
at small $x$ ($< 0.01$), and the determination of the NPDFs
is improved especially in this small-$x$ region. 

Because the NLO contributions are obvious only in the region, $x<0.01$,
it is very important to measure the $Q^2$ dependence to pin down
the NLO effects such as the gluon distributions. The possibilities
are measurements at future electron facilities such as eRHIC \cite{erhic}
and eLIC \cite{elic}.

\subsection{Parton distribution functions in nuclei}
\label{npdfs} 

\begin{figure}[b]
        \includegraphics*[width=80mm]{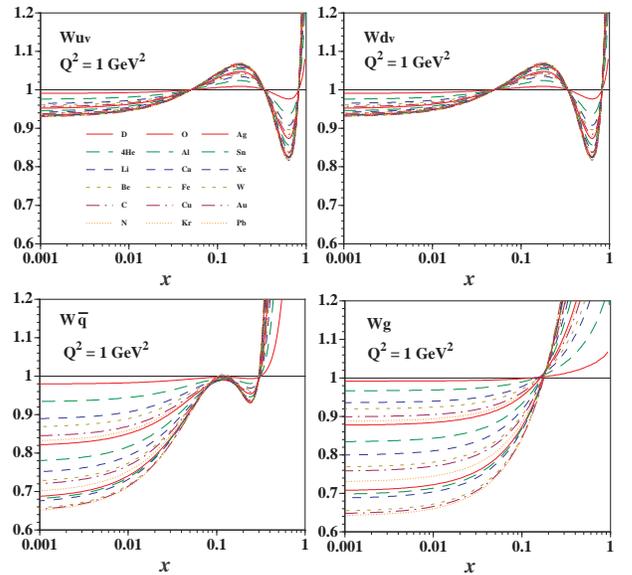} 
        \vspace{-0.2cm}
\caption{(Color online) 
Nuclear modifications $w_i$ ($i=u_v$, $d_v$, $\bar q$, and $g$) are shown
in the NLO for all the analyzed nuclei and $^{16}O$ at $Q^2$=1 GeV$^2$. 
As the mass number becomes larger in the order of D, $^4$He, Li, ..., and Pb,
the curves deviate from the line of unity ($w_i=1$).}
\label{fig:npdfs-all}
\end{figure}

Nuclear modifications of the PDFs are shown for all the analyzed
nuclei and $^{16}O$ at $Q^2$=1 GeV$^2$ in Fig. \ref{fig:npdfs-all}. 
It should be noted that the modifications of $u_v$ are the same
as the ones of $d_v$ in isoscalar nuclei, but they are different
in other nuclei. The modifications increase as the nucleus becomes
larger, and the dependence is controlled by the overall $1/A^{1/3}$
factor and the $A$ dependence in Eq. (\ref{eqn:more-a}).
The extreme values ($x_{0i}^+$, $x_{0i}^-$) are assumed to be
independent of $A$ in our current analysis as explained in
Sec. \ref{paramet}, so that they are the same in Fig. \ref{fig:npdfs-all}.
Although the oxygen data are not used in our global analysis, its PDFs
are shown in the figure because they are useful for an application to
neutrino oscillation experiments \cite{nuint}. Our code is supplied
at the web site in Ref. \cite{npdf-lib} for calculating the NPDFs
and their uncertainties at given $x$ and $Q^2$. 

\begin{figure}[t]
        \includegraphics*[width=40mm]{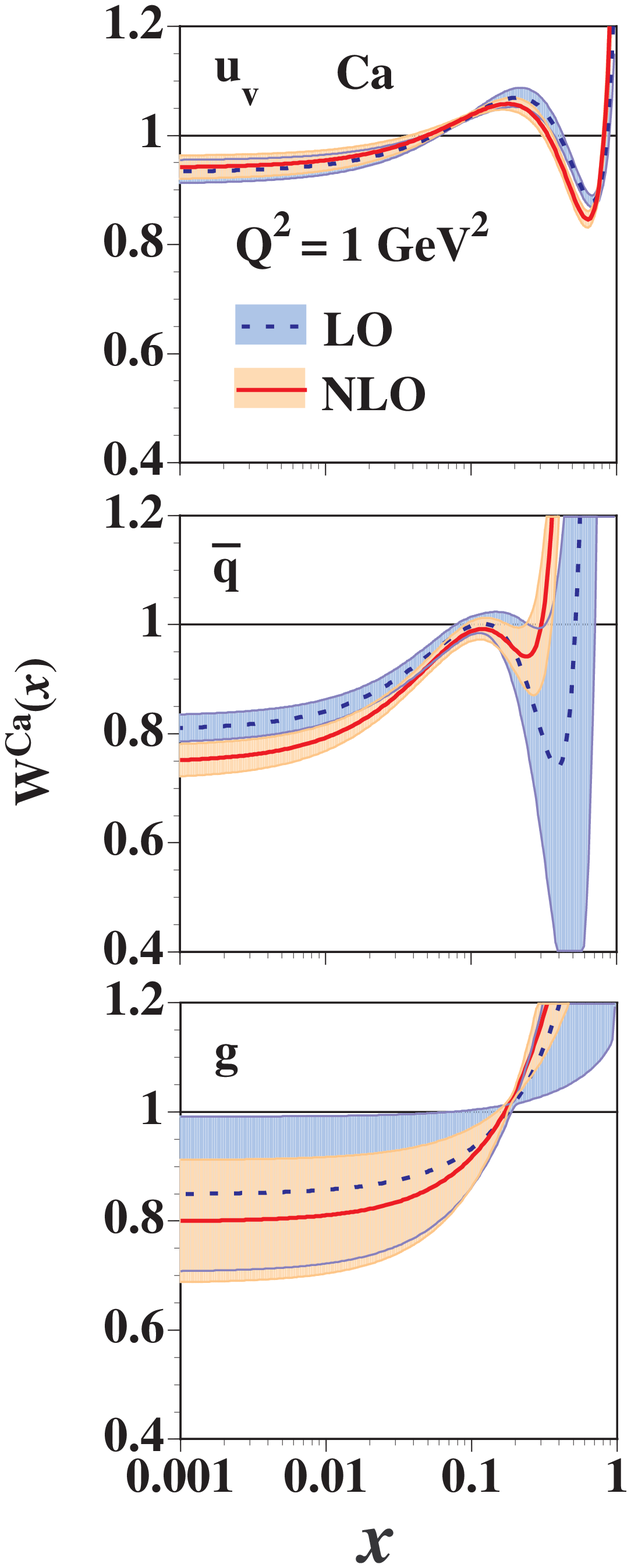} \hspace{0.5mm}
        \includegraphics*[width=40mm]{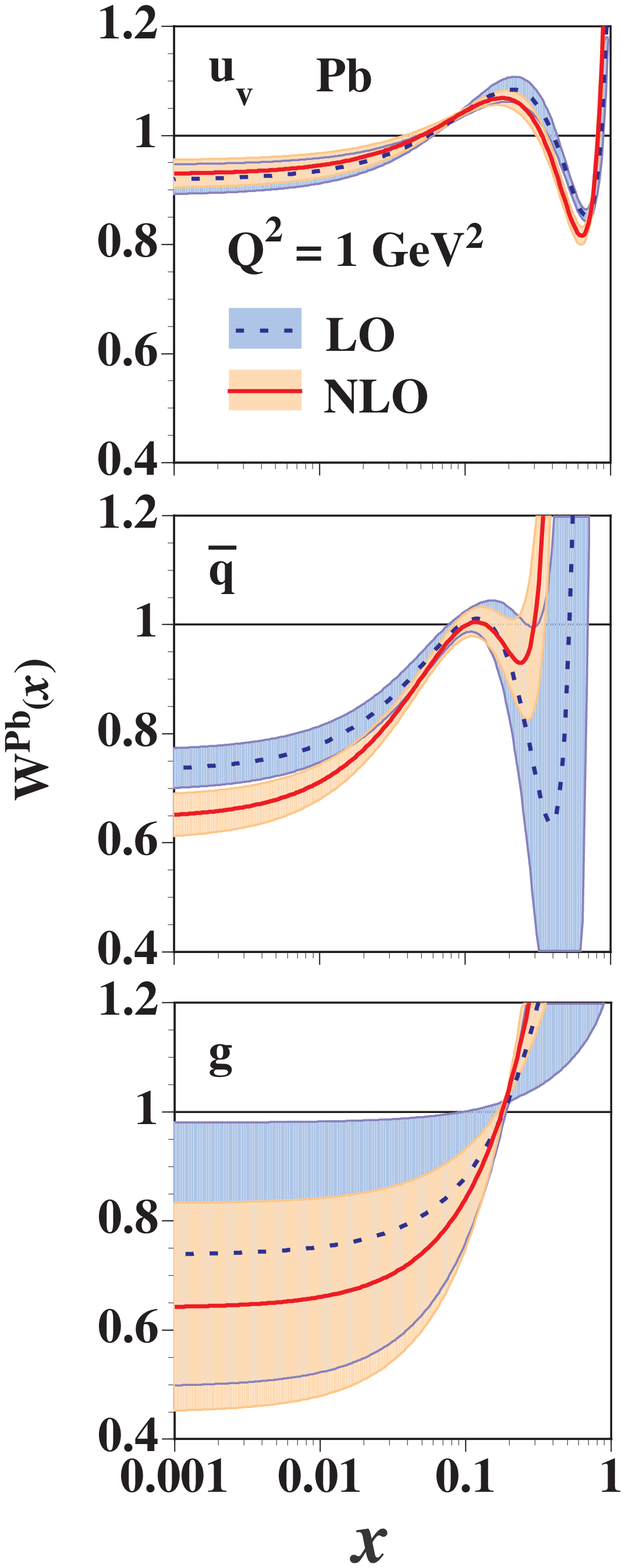} 
        \vspace{-0.2cm}
\caption{(Color online) 
Nuclear modifications of the PDFs and their uncertainties are shown
for the calcium and lead nuclei at $Q^2$=1 GeV$^2$.
    The dashed and solid curves indicate LO and NLO results, and
    LO and NLO uncertainties are shown by the dark- and
    light-shaded bands, respectively.}
\label{fig:npdfs-ca-pb}
\end{figure}

As examples of medium and large nuclei, we take the calcium and lead
and show their distributions and uncertainties at $Q^2$=1 GeV$^2$
in Fig. \ref{fig:npdfs-ca-pb}. Because the deuteron is a special nucleus
and it needs detailed explanations, its results are separately discussed 
in Sec. \ref{deuteron}. 
The figure indicates that valence-quark distributions
are determined well in the wide range, $0.001<x<1$ because the uncertainties
are small. It is also interesting to find that the LO and NLO uncertainties
are almost the same. There are following reasons for these results.
The valence-quark modifications at $x>0.3$ are determined by 
the accurate measurements of $F_2$ modifications. The antishadowing
part in the region, $0.1<x<0.2$, is also determined by the $F_2$ data
because there is almost no nuclear modification in the antiquark
distributions according to the Drell-Yan data. If the valence-quark
distributions are obtained at $x>0.1$, the small-$x$ behavior is
automatically constrained by the baryon-number and charge conservations
in Eq. (\ref{eqn:3conserv}).
Because of these strong constraints, the LO and NLO results are not
much different. In the near future, the JLab (Thomas Jefferson
National Accelerator Facility) measurements will provide data 
which could constraint the nuclear valence-quark
distribution especially at medium and large $x$ \cite{jlab-emc07}.
Furthermore, future neutrino measurements such as the MINER$\nu$A
experiment at Fermilab \cite{minerva} and the one at a possible
neutrino factory \cite{nu-fact} should provide important information
at small $x$ ($x=0.01 - 0.1$).

The antiquark distributions are also well determined except for
the large-$x$ region, $x>0.2$, because there is no accurate Drell-Yan
data and the structure functions $F_2^A$ are dominated by
the valence-quark distributions. The LO and NLO uncertainties are
similar except for the large-$x$ region. Because of gluon contributions
in the NLO, the antiquark shadowing modifications are slightly different
between the LO and NLO. Possible large-$x$ Drell-Yan measurements such as 
J-PARC (Japan Proton Accelerator Research Complex) \cite{jparc-dy}
and Fermilab-E906 \cite{e906} should be valuable for
the nuclear antiquark distributions in the whole-$x$
range. In the similar energy region, there is also the GSI-FAIR
(Gesellschaft f\"ur Schwerionenforschung 
 -Facility for Antiproton and Ion Research) project \cite{gsi-fair}.

The gluon distributions contribute to the $F_2$ and Drell-Yan
ratios as higher-order effects. Therefore, they should be
determined more accurately in the NLO analysis than the LO one.
Such tendencies are found in Fig. \ref{fig:npdfs-ca-pb}
because the NLO uncertainties are smaller than the LO ones
in both carbon and lead. However, these NLO improvements are
not as clear as the cases of polarized PDF \cite{aac0306}
and fragmentation functions \cite{hkns07}. It is because 
the $Q^2$-dependent data are not accurate enough to probe such
higher-order effects as discussed in Sec. \ref{q2-dependence}.
In order to fix the gluon distributions, accurate measurements
are needed for the scaling violation of $F_2^A/F_2^{A'}$
\cite{erhic,elic}. The gluon distributions should be also probed
by production processes of such as charged hadrons \cite{lw02},
heavy flavor \cite{kope2002}, $\Upsilon$ \cite{frankfurt-lhc03},
low-mass dilepton \cite{fai05}, and direct photon \cite{at06-gluon-photon}. 
There is a recent study on the gluon shadowing from the HERA diffraction
data \cite{arsence-07}. Nuclear gluon distributions play an important
role in discussing properties of quark-hadron matters 
in heavy-ion reactions, so that they need to be determined experimentally.

\begin{figure}[t]
        \includegraphics*[width=80mm]{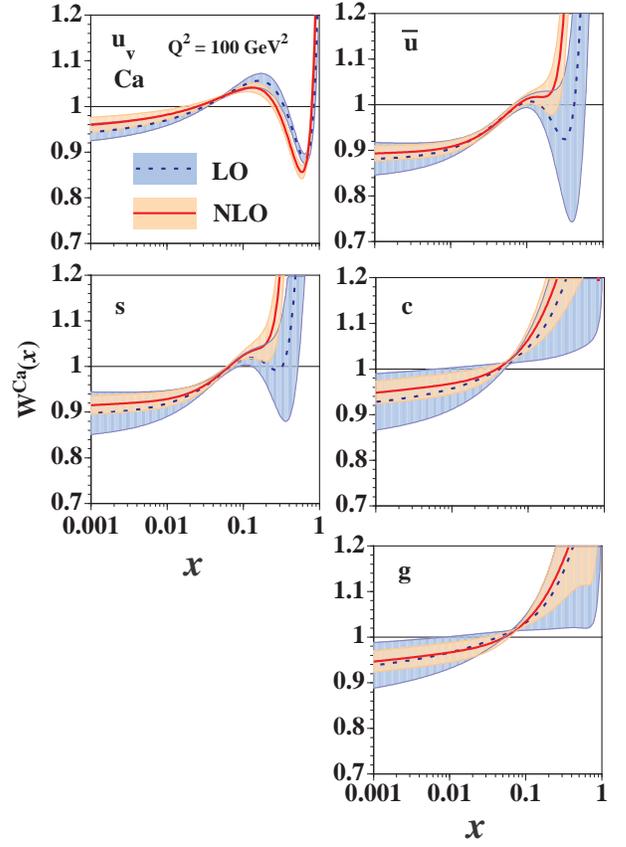} 
        \vspace{-0.2cm}
\caption{(Color online) 
Nuclear modifications are shown for the calcium at $Q^2$=100 GeV$^2$.}
\label{fig:w-ca-100}
\end{figure}

We also show the nuclear modifications at $Q^2$=100 GeV$^2$ for the calcium
in Fig. \ref{fig:w-ca-100}. The nuclear modifications are not very different
from those at $Q^2$=1 GeV$^2$ for the valence-quark distributions.
However, the shadowing corrections become smaller in the antiquark and
gluon distributions in comparison with the ones at $Q^2$=1 GeV$^2$,
and the modifications tend to increase at medium and large $x$.

\begin{figure}[t]
        \includegraphics*[width=73mm]{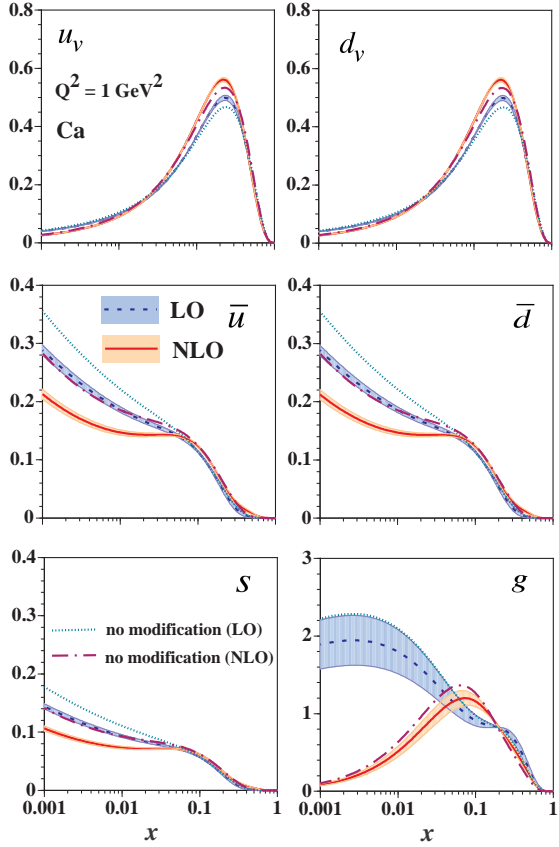} 
        \vspace{-0.2cm}
\caption{(Color online) 
Parton distribution functions in the calcium at $Q^2$=1 GeV$^2$.
``No modification" indicates, for example, the distribution $(Zu_v+Nd_v)/A$
in the figure of $u_v^A$.}
\label{fig:npdf-ca}
\end{figure}

\begin{figure}[t]
        \includegraphics*[width=73mm]{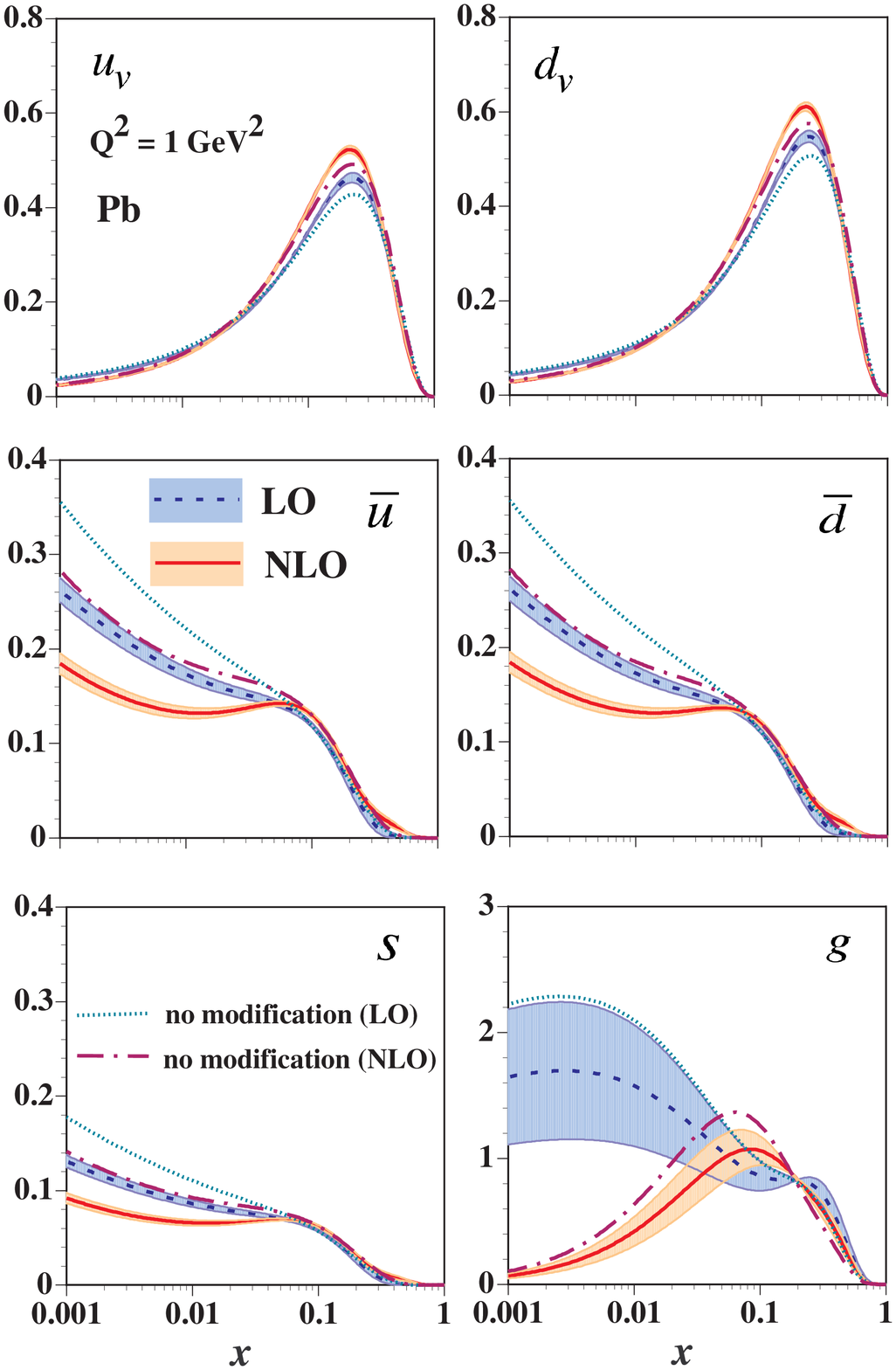} 
        \vspace{-0.2cm}
\caption{(Color online) 
Parton distribution functions in the lead at $Q^2$=1 GeV$^2$.
``No modification" indicates, for example, the distribution $(Zu_v+Nd_v)/A$
in the figure of $u_v^A$.}
\label{fig:npdf-pb}
\end{figure}

The distribution functions themselves and their uncertainties are
shown in Figs. \ref{fig:npdf-ca} and \ref{fig:npdf-pb} for the calcium
and lead, respectively. Both LO and NLO distributions are shown.
Here, the uncertainties from the nucleonic PDFs are not included
in the uncertainty bands. The calcium is an isoscalar nucleus,
so that $u_v^A$ and $\bar u^A$ are equal to $d_v^A$ and $\bar d^A$.
However, they are different in the lead nucleus because of the neutron
excess. In order to see nuclear modification effects, we show
the distributions without the nuclear modifications. For example,
the distribution $(Zu_v+Nd_v)/A$ is shown in the figure of $u_v^A$
by using the MRST distributions for $u_v$ and $d_v$. Although
the uncertainties are large in the antiquark and gluon distributions
at medium and large $x$, they are not very conspicuous in 
Figs. \ref{fig:npdf-ca} and \ref{fig:npdf-pb} because
the distributions themselves are small.

\begin{figure}[h!]
        \includegraphics*[width=60mm]{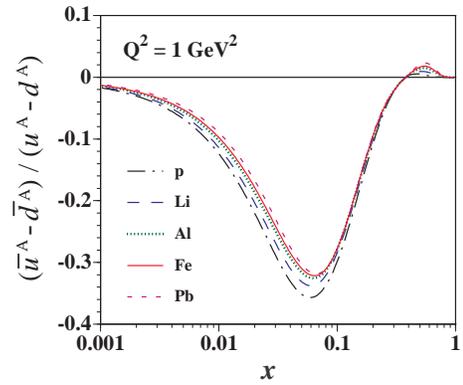} 
        \vspace{-0.2cm}
\caption{(Color online) 
The ratio of flavor asymmetric distributions,
$(\bar u^A -\bar d^A)/(u^A-d^A)$, is shown for the proton, lithium,
aluminum, iron, and lead at $Q^2$=1 GeV$^2$. In the isoscalar nuclei,
the distributions vanish ($\bar u^A -\bar d^A=0$).}
\label{fig:ub-db}
\end{figure}

The flavor asymmetric antiquark distributions are assumed in this analysis
as they are defined in Eq. (\ref{eqn:wpart}). From this definition,
it is obvious that $\bar u^A$ and $\bar d^A$ are equal in isoscalar
nuclei such as carbon and calcium. In Fig. \ref{fig:ub-db}, the ratio 
$(\bar u^A -\bar d^A)/(u^A-d^A)$ is shown for the proton (p), lithium (Li),
aluminum (Al), iron (Fe), and lead (Pb) at $Q^2$=1 GeV$^2$.
Because the nuclear corrections are assumed to be equal for the antiquark
distributions at $Q_0^2$ in Eq. (\ref{eqn:wpart}), they are almost
independent of nuclear species except for the isoscalar nuclei.
It is interesting to investigate possible nuclear modifications
on the distribution $\bar u-\bar d$ at the future facilities 
\cite{jparc-dy,e906}.

There are noticeable differences between our NLO analysis results
and the ones in Ref. \cite{ds04} especially in the strange-quark
and gluon modifications. These differences come from various
sources. First, the analyzed experimental data sets are slightly
different. Second, the strange-quark distributions are created
by the DGLAP evolution by assuming $s(x)=0$ at the initial $Q^2$ scale,
and the charm distributions are neglected in Ref. \cite{ds04}.
These differences lead to the discrepancies of the gluon modifications.

The determined NPDFs and their uncertainties can be calculated by
using our code, which is supplied on our web site \cite{npdf-lib}.
By providing a kinematical condition for $x$ and $Q^2$ and also
a nuclear species, one can calculate the NPDFs. It is explained
in Appendix \ref{library}.
If one needs analytical expressions of the NPDFs at the initial
scale $Q_0^2$, one may read instructions in Appendix \ref{appen-a}.

\section{Nuclear modifications in deuteron}
\label{deuteron} 

Nuclear densities are usually independent of the mass number, which
indicates that the average nucleon separation is constant in nuclei.
However, the deuteron is a special nucleus in the sense that its
radius is about 4 fm, which is much larger than the average nucleon
separation in ordinary nuclei ($\sim 2$ fm).  Because it is
a dilute system, nuclear modifications are often neglected.
In fact, corrections to nucleonic structure functions and PDFs are small,
namely within a few percentages according to theoretical estimates
\cite{bk-shadow-1994, d-zn-92, d-bgnpz-93, d-mst-93, d-prw-95}
even if they are taken into account.

\begin{table}[b]
\caption{Theoretical model estimations for nuclear modification 
         of $F_2$ in the deuteron at $x \sim 0.01$.}
\label{tab:d-modification}
\begin{ruledtabular}
\begin{tabular*}{\hsize}
{c@{\extracolsep{0ptplus1fil}}c@{\extracolsep{0ptplus1fil}}c
@{\extracolsep{0ptplus1fil}}c}
Reference  & $Q^2$ (GeV$^2$) & $x$  & Modification (\%)  \\
\colrule
\cite{bk-shadow-1994}  &  4    &  0.015    &  1.1          \\
\cite{d-zn-92}         &  4    &  0.010    &  2.5          \\
\cite{d-bgnpz-93}      &  4    &  0.010    &  2.0          \\
\cite{d-mst-93}        &  4    &  0.010    &  0.4$-$0.8    \\
\cite{d-prw-95}        &  4    &  0.010    &  1$-$2 (3.5)  \\
\end{tabular*}
\end{ruledtabular}
\end{table}
\vspace{+0.0cm}

Theoretical modifications in the deuteron depend
much on models. The shadowing of Ref. \cite{bk-shadow-1994},
which was used in the MRST analysis, is 1.1\% at $x$=0.015 and
$Q^2$=4 GeV$^2$; however, other model calculations are different,
0.5$-$3.5\% at $x\sim 0.01$ as shown in Table \ref{tab:d-modification}.
Furthermore, the modifications at medium $x$ could be as large as
or larger than the small-$x$ shadowing (for example, 
5\% at $x=0.6 - 0.7$ according to Ref. \cite{d-mst-93}).

Such deuteron corrections are becoming important recently
although the magnitude itself may not be large. For example, precise nuclear
modifications need to be taken into account for investigating quark-hadron
matters in heavy-ion collisions such as deuteron-gold reactions in
comparison with deuteron-deuteron ones at RHIC \cite{rhic-heavy-ion}.
They are also valuable for discussing Gottfried-sum-rule violation and
flavor-asymmetric antiquark distributions because deuteron targets
are used \cite{flavor3}.
 
In our previous versions on the NPDFs \cite{hkm01,hkn04}, the data
of $F_2^D/F_2^p$ are not included in the used data set. Obtained
nuclear modifications tend to be large in comparison with
the theoretical model estimations.
For example, the HKN04 analysis \cite{hkn04,npdf-lib}
indicates about 6 and 8\% corrections in 
antiquark and gluon distributions, respectively, at small $x$
($\sim 0.01$) with $Q^2$=1 GeV$^2$ and about 5\% in valence-quark ones
at medium $x$ ($\sim 0.7$). They are possibly overestimations
in the sense that typical theoretical models have corrections
within the order of a few percentages.

In order to obtain reasonable deuteron modifications from experimental
data, we added $F_2^D/F_2^p$ measurements into the data set
in our $\chi^2$ analysis. These deuteron data were already included
in the analysis results in Sec. \ref{results}. In addition to
the analysis of Sec. \ref{results}, two other analyses have been
made by modifying Eq. (\ref{eqn:wi}):
\begin{equation}
w_i(x,A,Z)=1+ e_d \left( 1 - \frac{1}{A^\alpha} \right) 
          \frac{a_i +b_i x+c_i x^2 +d_i x^3}{(1-x)^{\beta_i}}.
\label{eqn:wi-d}
\end{equation}
An additional factor $e_d$ is introduced. The analysis results
in Sec. \ref{results} correspond to the $e_d$=1 case.
The other analyses have been made by taking $e_d$=0 and
by taking it as a free parameter. We call them analyses 1, 2, and 3,
respectively:
\begin{itemize}
\item analysis 1: $e_d$=1,
\item analysis 2: $e_d$=0,
\item analysis 3: $e_d$=free parameter.
\end{itemize}
The $\chi^2$ values of these analyses are listed in Table \ref{tab:chi2-d}.

\begin{table}[b]
\caption{Each $\chi^2$ contribution.}
\label{tab:chi2-d}
\begin{ruledtabular}
\begin{tabular*}{\hsize}
{c@{\extracolsep{0ptplus1fil}}c@{\extracolsep{0ptplus1fil}}c
@{\extracolsep{0ptplus1fil}}c@{\extracolsep{0ptplus1fil}}c}
Nucleus  & \# of data & $\chi^2$ ($e_d=1$) 
         & $\chi^2$ ($e_d=0$) & $\chi^2$ (free) \\
\colrule
D/p      &      290    &       322.5   &   282.6     &   284.9   \\
$F_2^A/F_2^D$
         &      606    &       709.3   &   704.8     &   703.8   \\
$F_2^{A_1}/F_2^{A_2}$
         &      293    &       369.0   &   381.6     &   375.3   \\
Drell-Yan
         &   \   52    &  \     85.1   & \  84.5     & \  85.4   \\
\colrule\colrule
Total    &     1241    &      1485.9 \ &  1453.4 \   & 1449.4 \  \\
($\chi^2$/d.o.f.)  &   &      (1.21)   &  (1.18)     & (1.18)    \\
\end{tabular*}
\end{ruledtabular}
\end{table}
\vspace{+0.0cm}

\begin{figure}[t]
        \includegraphics*[width=80mm]{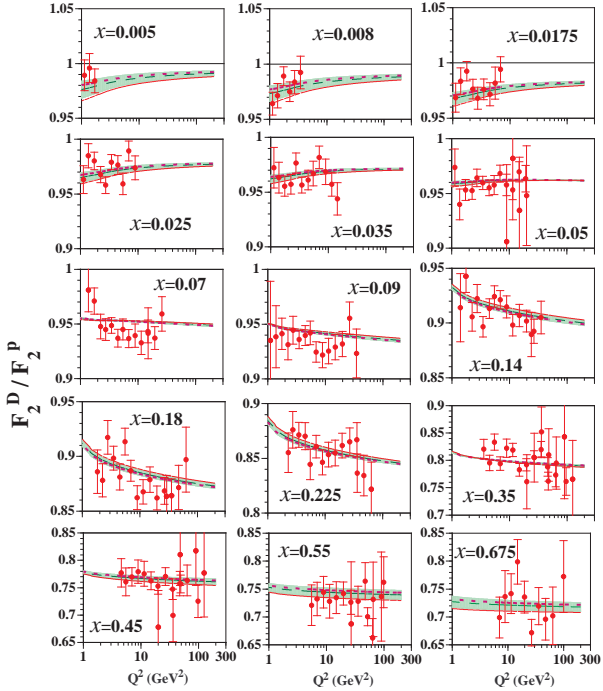} 
        \vspace{-0.2cm}
\caption{(Color online) Comparison with the NMC data of
    $F_2^{D}/F_2^{p}$. The solid, dotted, and dashed curves
    indicate the NLO results of the analyses 1 ($e_d=1$), 2 ($e_d=0$),
    and 3 ($e_d$=free), respectively. The shaded bands indicate uncertainties
    of the analysis-3 curves.}
\label{fig:f2dp}
\end{figure}

The deuteron modifications are terminated in the analysis 2 by taking
$e_d=0$. Although such an assumption does not seem to make sense, we
found a smaller $\chi^2$ value from the $F_2^D/F_2^p$ data than
the one of the first analysis
as shown in Table \ref{tab:chi2-d}. There are three major reasons
for this result. First, the deuteron modifications obtained
by the overall $A$ dependence in Eq. (\ref{eqn:wi}) are too large
because the deuteron is a loosely bound system which is much different
from other nuclei. Smaller modifications are expected from the
large nucleon separation. 
Second, deuteron data are used for determining the PDFs in the ``nucleon"
\cite{mrst98} by considering nuclear shadowing modifications of 
Ref. \cite{bk-shadow-1994}. The modifications are calculated in
a vector-meson-dominance mechanism and the shadowing in the deuteron
is about 1\% at $x \sim 0.01$ according to this model. If this shadowing
is not a realistic correction, the nucleonic PDFs of the MRST should
partially contain nuclear effects at small $x$.
The medium- and large-$x$ regions are not corrected, so that some
deuteron effects could be also included in the nucleonic PDFs
in these regions. However, the corrections are not experimentally
obvious in such $x$ regions as we find in the actual data of 
Fig. \ref{fig:f2dp}. In this way, the nucleonic PDFs could contain
some deuteron modification effects. 
Third, the nuclear effects could be absorbed into the $\bar u/\bar d$
asymmetry because it is determined partially by the $F_2^D/F_2^p$ data.
These are the possible reasons why analysis 2 produces
the smaller $\chi^2$ value. 

In the third analysis, the additional parameter $e_d$ is determined
from the global analysis. As mentioned, the internucleon separation
is exceptionally large in the deuteron. It leads
to small nuclear corrections, which are much smaller than a smooth
$A$-dependent functional form, as calculated in various models
\cite{bk-shadow-1994, d-zn-92, d-bgnpz-93, d-mst-93, d-prw-95}
in Table \ref{tab:d-modification}.
These models could be used for estimating an appropriate value for $e_d$.
However, we try to determine the nuclear PDFs without
relying on specific theoretical models.
The modification parameter $e_d$ is determined from
the experimental data, and our analysis 3 indicates
$e_d=0.304\pm 0.141$. If the diffuse deuteron system is considered,
the 70\% reduction may make sense. Nonetheless, it should be
noted that this factor may not reflect a realistic modification
because it is likely that the deuteron effects are contained
in the nucleonic PDFs. 

We show actual comparisons with the experimental data for $F_2^D/F_2^p$
by the NMC in Fig. \ref{fig:f2dp}. Three global analysis results are
shown, and the shaded areas indicate uncertainty bands in analysis 3.
The figure suggests that all the analyses are
successful in explaining the data. However, as indicated in the $\chi^2$
reductions in the analyses 2 and 3, it is clear that their curves are
closer to the experimental data at small $x$ such as $x$=0.005 and 0.008.
There is a tendency that deviations from the data become larger as
the deuteron modifications are increased. All the analysis results are
more or less similar in the medium- and large-$x$ regions. All the curves
of the analysis 1 ($e_d=1$) and 2 ($e_d=0$) are within the uncertainty
bands of the analysis 3 ($e_d$=free), although the analysis-1 curves
are at the edges of the error bands at small $x$. It means that all these
analysis are consistent with each other and they explain the experimental
data.

\begin{figure}[t]
        \includegraphics*[width=80mm]{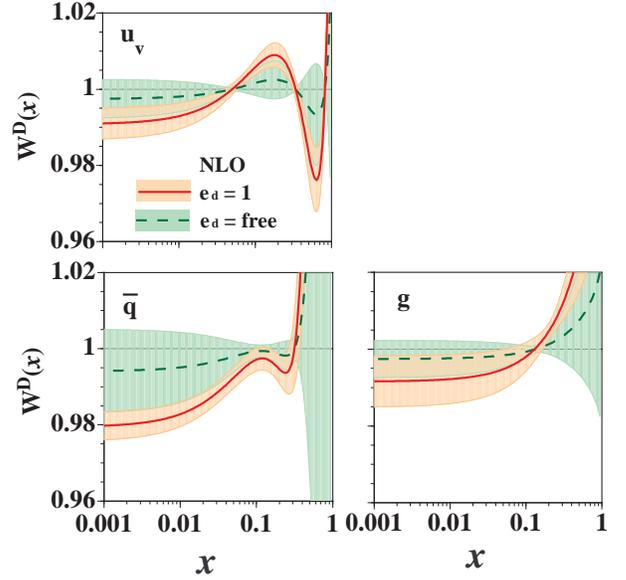}
        \vspace{-0.2cm}
\caption{(Color online) 
Nuclear modifications of the PDFs and their uncertainties are shown
for the deuteron at $Q^2$=1 GeV$^2$. The solid and dashed curves
are obtained by analyses 1 and 3, respectively, and their
uncertainties are shown by the shaded bands.}
\label{fig:npdfs-d}
\end{figure}

Modifications of the PDFs are shown for the deuteron in Fig. \ref{fig:npdfs-d}
at $Q^2$=1 GeV$^2$. The results of analyses 1 and 3 are shown with
their uncertainty estimation. There is no deuteron modification
in analysis 2 as obvious from the definition in Eq. (\ref{eqn:wi-d}).
The uncertainty bands of analysis 3 shrink at the points, where
the nuclear modifications vanish ($w^D=1$),  for example,
in the figure of the valence-quark modification $w_{u_v}$.
This is caused by the error from the parameter $e_d$. 
Its contributions to the uncertainties are large and the derivative
$\partial w_{u_v}/\partial e_d$ is proportional to
$a_{u_v}+b_v x +c_v x^2+d_v x^3$, which vanishes at the same
points as the function $w_{u_v}^D(x)-1=0$ in Eq. (\ref{eqn:wi-d}).
It leads to the gourd-shaped uncertainty band in $w_{u_v}$ because other
terms are small. 
Such an error shape does not appear in analysis 1 because
the error term of $e_d$ does not exist. Here, the derivative 
$\partial w_{u_v}/\partial d_v^{(1)}$ is also a cubic polynomial and
vanishes at three $x$ points. However, $\partial w_{u_v}/\partial x_{0v}^+$
and $\partial w_{u_v}/\partial x_{0v}^-$ are quadratic functions,
which do not vanish at the same $x$ points, and their contributions
to the uncertainties are of the same order of the $d_v^{(1)}$
and $d_v^{(2)}$ terms. This is the reason why such a gourd-shaped
function does not appear in the uncertainties of the $e_d=1$ analysis.

The antiquark shadowing is about 2\% at $x=0.01$ and the valence-quark
modification is about 1\% in the analysis 1 according to
Fig. \ref{fig:npdfs-d}. We should note that the antiquark shadowing
is reduced about 30\% at $Q^2$=4 GeV$^2$ as shown in Fig. \ref{fig:cad-q2}
in comparing it with the theoretical values in
Table \ref{tab:d-modification}.
In analysis 3, the antiquark shadowing becomes 0.5\%,
which could be slightly smaller than the theoretical ones. 
Both corrections (2\% and 0.5\%) are slightly different from
the assumed correction (1\% at $x$=0.01 and $Q^2$=4 GeV$^2$) in the MRST fit.
The uncertainty bands depend on the initial functional form or assignment
of the parameters in the $\chi^2$ analysis. The uncertainties are generally
larger in analysis 3, and the line of $w^D=1$ is within the bands.
It suggests that precise deuteron modifications cannot be determined
at this stage. The modifications in Fig. \ref{fig:npdfs-d} may not be
seriously taken because the deuteron effects could be partially included
in the nucleonic PDFs. It is difficult to judge what the realistic deuteron
modifications are at this stage. Actual modifications are possibly
in between these analysis results.

In order to obtain realistic modifications, the nucleonic PDFs should
be determined by considering the deuteron modifications, for example,
of our analysis results. Then, using new nucleonic PDFs, we redetermine
our nuclear PDFs including the ones in the deuteron. Realistic
deuteron modifications should be obtained by repeating this step.

\section{\label{summary} Summary}

Nuclear PDFs have been determined by the global analyses of experimental
data for the ratios of the $F_2$ structure functions and Drell-Yan cross
sections. The uncertainties of the determined NPDFs are estimated by
the Hessian method. The first important point is that the uncertainties
were obtained in both LO and NLO so that we can discuss the NLO
improvement on the determination. We found slight NLO improvements
for the antiquark and gluon distributions at small $x$ ($\sim 0.001$);
however, they are not significant at larger $x$.
Accurate experimental measurements, especially on the $Q^2$ dependence
at small $x$, should be useful for determining higher-order effects
such as the nuclear gluon distributions. 

The valence-quark distributions are well determined.
The antiquark distributions are also determined at $x<0.1$; however,
they have large uncertainties at $x>0.2$. The gluon modifications
are not precisely determined in the whole $x$ region.
Future measurements are needed to determine accurate nuclear
distributions.

Nuclear modifications were discussed for the deuteron in comparison
with the experimental data of $F_2^D/F_2^p$.
However, it is difficult to find accurate
modifications at this stage because deuteron effects could be partially
contained in the nucleonic PDFs. An appropriate nucleonic PDF analysis
is needed in addition to accurate measurements on the ratio $F_2^D/F_2^p$.

Our NPDFs and their uncertainties can be calculated by using the codes
in Ref. \cite{npdf-lib}.

\begin{acknowledgements}
M.H. and S.K. were supported by the Grant-in-Aid for Scientific
Research from the Japanese Ministry of Education, Culture, Sports,
Science, and Technology. T.-H.N. was supported by Japan Society
for the Promotion of Science.
\end{acknowledgements}

\appendix
\section{\label{appen-a} Parameters $a_{u_v}$, $a_{d_v}$, and $a_g$}

The determined parameters are listed in Table \ref{table:parameters}.
There are other parameters, $a_{u_v}$, $a_{d_v}$, and $a_g$,
which are automatically calculated from the tabulated values
by the conservation conditions of nuclear charge, baryon number,
and momentum in Eq. (\ref{eqn:3conserv}). Because they depend on
nuclear species, namely on $A$ and $Z$, their calculation method
is explained in the following.
The flavor asymmetric antiquark distributions in Eq. (\ref{eqn:wpart})
are used in this analysis, whereas the flavor symmetric ones
are used in the previous versions in Refs. \cite{hkm01,hkn04}, so that
relations are slightly different from the ones in Appendix of
Ref. \cite{hkm01}. One can obtain values of these parameters
for any nuclei by calculating the following integrals:
\begin{alignat}{2}
\! \! \! \! \! \! \!  
  I_1(A) & = \int dx \frac{H_v(x,A)}{(1-x)^{\beta_v}}u_v(x),
& 
\! \! \! \! \! \! \! \! \! \! \! \! 
I_2(A) & = \int dx \frac{H_v(x,A)}{(1-x)^{\beta_v}}d_v(x),
\nonumber \\
  I_3      & = \int dx \frac{1}{(1-x)^{\beta_v}}u_v(x),
& I_4      & = \int dx \frac{1}{(1-x)^{\beta_v}}d_v(x),
\nonumber \\
  I_5      &= \int dx \frac{x}{(1-x)^{\beta_v}}u_v(x),
& I_6      &= \int dx \frac{x}{(1-x)^{\beta_v}}d_v(x),
\nonumber \\
\! \! \! \! \! \! \! 
I_7(A) & = \int dx x \biggl[ \frac{H_v (x,A)} {(1-x)^{\beta_v}}
                              \{u_v(x)+ d_v(x)\} 
\! \! \! \! \! \! \! \! \! \! \! \! \! \! \! \! \! \!  
                              & 
\nonumber \\
& \! \! \! \! \! \! \! \! \! \! \! \! \! \! \! 
+ \frac{a_{\bar q}(A)+H_{\bar q}(x,A)}{(1-x)^{\beta_{\bar q}}}
   2 \{ \bar u(x)+\bar d(x)+\bar s(x) \}
   +\frac{H_g(x,A)}{(1-x)^{\beta_g}}g(x) \biggr] , 
\! \! \! \! \! \! \! \! \! \! \! \! \! \! \! \! \! \! \! \! \! \! 
\! \! \! \! \! \! \! \! \! \! \! \! \! \! \! \! \! \! \! \! \! \!    
\! \! \! \! \! \! \! \! \! \! \! \! \! \! \! \! \! \! \! \! \! \! 
\! \! \! \! \! \! \! \! \! \! \! \! \! \! \! \! \! \! \! \! \! \!    
&
\nonumber \\
I_8       &= \int dx \frac{x}{(1-x)^{\beta_g} }g(x), & 
\end{alignat}
where $\beta_v=\beta_{\bar q}=\beta_g=0.1$ and
$H_i(x,A)$ is 
\begin{equation}
H_i(x,A)=b_i x+c_i x^2+d_i(A) x^3.
\end{equation}
Here, the parameters $d_v$, $a_{\bar q}$, $d_{\bar q}$, and $d_{g}$
depend on $A$ according to Eq. (\ref{eqn:more-a}). It is noteworthy that
the integrals $I_1$, $I_2$, and $I_7$ depend on $A$.
Using these integrals, one obtains parameter values 
by the conservations in Eq. (\ref{eqn:3conserv}):
\begin{equation}
a_{u_v}(A,Z) = - \frac{Z I_1(A) + (A-Z) I_2(A)}{Z I_3 + (A-Z) I_4} ,
\label{eqn:auv}
\end{equation}
\begin{equation}
a_{d_v}(A,Z) = - \frac{Z I_2(A) + (A-Z) I_1(A)}{Z I_4 + (A-Z) I_3} ,
\label{eqn:adv}
\end{equation}
\begin{align}
a_{g}(A,Z)   & = - \frac{1}{I_8} \,  \bigg[ \,
            a_{u_v}(A,Z) \left\{ \frac{Z}{A} I_5
                                +\left(1-\frac{Z}{A} \right) I_6 \right\}
\nonumber \\
         & \! \! \! \! \! \! \! \! \! \! \! \! \! \! \! 
           +a_{d_v}(A,Z) \left\{ \frac{Z}{A} I_6
                                +\left(1-\frac{Z}{A} \right) I_5 \right\}
           +I_7(A) \, \bigg]   .
\label{eqn:ag}
\end{align}

From these values together with the parameters in 
Table \ref{table:parameters} and the nucleonic PDFs at $Q^2$=1 GeV$^2$
of the MRST parametrization \cite{mrst98}, 
one obtains the NPDFs at $Q^2$=1 GeV$^2$
for a given nucleus. If distributions are needed at different $Q^2$,
one needs to evolve the NPDFs by using own evolution code.
If one does not have such an evolution code, one had better use
the code in Appendix B. Numerical values of the NPDFs can be
obtained for given $x$, $Q^2$, and $A$. In particular, if one is
interested in estimating the uncertainties of the NPDFs, this practical
code needs to be used.

\section{\label{library} Code for calculating nuclear
         parton distribution functions and their uncertainties}

Codes for calculating nuclear PDFs and their uncertainties can be
obtained from the web site \cite{npdf-lib}. A general guideline for
usage is explained in Appendix B of Ref. \cite{hkn04}, and the conditions
are the same in the current version, HKN07 (Hirai, Kumano, Nagai in 2007).
Therefore, the details should be found in Ref. \cite{hkn04}.
The NPDFs can be calculated for nuclei at given $x$ and $Q^2$,
for which the kinematical ranges should be in $10^{-9}\le x \le 1$
and  1 GeV$^2 \le Q^2 \le 10^8$ GeV$^2$. 

Input parameters for running the code are explained in the beginning
of the file npdf07.f. A sample program, sample.f, is supplied as
an example for calculating the nuclear PDFs and uncertainties.
The uncertainties of the NPDFs are estimated by using the Hessian
matrix and grid data for derivatives with respect to the parameters.
The usage is explained in the sample program.


\end{document}